\documentclass[12pt,a4paper]{article}
\usepackage{amsfonts}
\usepackage{amsmath}
\usepackage{amssymb}
\usepackage{latexsym}
\numberwithin{equation}{section}

\newcommand{\lanln}[1]{$\langle$\texttt{arXiv:#1}$\rangle$}
\newcommand{\SLtwor}{{\mathrm{SL}}(2,\mathbb{R})}
\newcommand{\sltwor}{{\mathfrak{sl}}(2,\mathbb{R})}
\newcommand{\BbbR}{\mathbb{R}}
\newcommand{\BbbZ}{\mathbb{Z}}
\newcommand{\be}{\begin{eqnarray}}
\newcommand{\ee}{\end{eqnarray}}
\newcommand{\nn}{\nonumber}
\newcommand{\com}[2]{\mbox{$[#1,#2]$}}

\newtheorem{lemma}{Lemma}[section]
\newtheorem{proposition}{Proposition}[section]
\newtheorem{theorem}[proposition]{Theorem}

\newcommand{\myproof}{\emph{Proof\/}. }
\newcommand{\myremark}{\emph{Remark\/}. }

\newcommand{\Aclass}{{\mathcal{A}_\mathrm{class}}}
\newcommand{\Vphys}{{\mathcal{V}_\mathrm{phys}}}
\newcommand{\Aphystar}{\mathcal{A}^{(\star)}_\mathrm{phys}}

\newcommand{\Haux}{{\mathcal{H}_\mathrm{aux}}}
\newcommand{\Hauxsym}{{\mathcal{H}_\mathrm{aux}^\mathrm{s}}}
\newcommand{\Hraq}{{\mathcal{H}_\mathrm{RAQ}}}

\newcommand{\Agrouphat}{{\hat{\mathcal{A}}_G}}

\newcommand{\qnought}{\textsf{q}_0}

\newcommand{\barGamma}{{\overline\Gamma}}
\newcommand{\barGammanought}{\barGamma_0}
\newcommand{\barGammareg}{\barGamma_{\mathrm{reg}}}
\newcommand{\barGammaex}{\barGamma_{\mathrm{ex}}}

\newcommand{\Mred}{\mathcal{M}}
\newcommand{\Mnought}{\mathcal{M}_0}
\newcommand{\Mreg}{\mathcal{M}_{\mathrm{reg}}}
\newcommand{\Mex}{\mathcal{M}_{\mathrm{ex}}}

\title{Group averaging in the $(p,q)$ oscillator
representation of $\SLtwor$}

\author{Jorma Louko\thanks{jorma.louko@nottingham.ac.uk}
\ and
Alberto Molgado\thanks{pmxam@nottingham.ac.uk}
\\
\noalign{\vspace{3ex}}
\small{\it School of Mathematical Sciences,
University of Nottingham,}\\
\small{\it Nottingham NG7 2RD, UK}
\\
\noalign{\vspace{1ex}}\\
\small{(Revised January 2004)} 
\\
\noalign{\vspace{1ex}}
\small{\lanln{gr-qc/0312014}}}

\date{}

\begin{document}


\maketitle

\begin{abstract}
We investigate refined algebraic quantisation with group averaging in
a finite-dimensional constrained Hamiltonian system that provides a
simplified model of general relativity. The classical theory has gauge
group $\SLtwor$ and a distinguished $\mathfrak{o}(p,q)$
observable algebra. The gauge group of the quantum theory is the
double cover of $\SLtwor$, and its representation 
on the auxiliary Hilbert space is isomorphic to the 
$(p,q)$ oscillator representation. When $p\ge2$, $q\ge2$ and $p+q
\equiv 0 \pmod2$, we obtain a 
physical Hilbert space with a nontrivial representation of the
$\mathfrak{o}(p,q)$ quantum observable algebra. For $p=q=1$, the
system provides the first example known to us where group averaging
converges to an indefinite sesquilinear form.
\end{abstract}

\newpage

\section{Introduction}

In quantisation of constrained systems, an elegant proposal to obtain
a physical inner product is to average
unconstrained quantum states in an auxiliary Hilbert space 
over the gauge group
\cite{AH,KL,QORD,epistle,BC,lands-against,lands-wren,%
GM2,GoMa,LouRov,Giulini-rev,Marolf-MG,Shvedov}.
When the averaging is formulated within refined algebraic
quantisation \cite{epistle,GM2,Marolf-MG} 
and converges in a sufficiently strong
sense, it provides 
either the unique rigging map, and hence the unique inner product on
the states that satisfy the constraints, or a
proof that the system does not admit a rigging map~\cite{GM2}.
Given the equivalence of refined algebraic quantisation to a wide class of
methods of choosing the physical inner product~\cite{Shvedov,GM1}, 
group averaging
thus provides considerable control over the quantisation. 

When the gauge group is compact, the averaging necessarily converges. 
For
a noncompact gauge group the averaging need not converge on all of
the auxiliary Hilbert space $\Haux$ but may still converge on a suitable
dense linear subspace~$\Phi$, and this is sufficient for
recovering the physical Hilbert space~$\Hraq$. The choice of the test
space $\Phi$ thus has a mathematical role in ensuring convergence, but it
also has a deep physical role in that $\Phi$ determines the algebra of
operators represented on $\Hraq$~\cite{GM2,GM1}. While
quantisation with group averaging can be carried out without the explicit
construction of any physical observables, 
in concrete examples one may
wish to choose $\Phi$ so that certain explicitly-known physical observables
of interest are contained in the algebra represented on~$\Hraq$. 

In this paper we study a quantum mechanical system whose constraints
mimic the Hamiltonian structure of general
relativity~\cite{DeWitt}. The constraint set consists of two
``Hamiltonian''-type constraints, quadratic in the momenta, and one
``momentum''-type constraint, linear in the momenta, and the classical
gauge group generated by these constraints is $\SLtwor$. The unreduced
phase space is $T^*\BbbR^{p+q} \simeq \BbbR^{2(p+q)}$, where $p\ge1$
and $q\ge1$. The system was introduced by Montesinos, Rovelli and
Thiemann with $p=q=2$~\cite{MRT}, and its quantisation with $p=q=2$
was studied in \cite{LouRov,MRT,Monte,GamPor} within Ashtekar's algebraic
quantisation~\cite{Ash1,Ash2}, in \cite{TruSL} within algebraic
constraint quantisation~\cite{trunk-kepler,trunk-pseudorigid}, in
\cite{LouRov,bojo-etal,bojo-strobl,kastrup} within group theoretic
quantisation~\cite{isham-les,GuiSte}, 
and in \cite{LouRov} within refined
algebraic quantisation with group averaging~\cite{epistle,GM2}.  All
these quantisations relied in one way or another on a distinguished
classical $\mathfrak{o}(2,2)$ observable algebra, constructing a
quantum theory in which these observables are promoted into quantum
operators. Within group averaging~\cite{LouRov}, it was in particular
found that a judicious choice for the test space is necessary to
achieve both convergence of the averaging and the inclusion of the
$\mathfrak{o}(2,2)$ observables in the physical operator algebra. 

For $p>2$ and $q=2$, the system has been studied in the context
of a ``two-time'' physical interpretation in
\cite{BarsKounnas,Bars98,Bars00,Bars01}. 
The case $p=q=2$ is currently being studied \cite{Dittrich-pc} within
the master constraint programme~\cite{Thiemann-master}. 
Related systems with $\SLtwor$ gauge invariance have been studied in 
\cite{SchallerStrobl,Tuyn,DEG}. 

We wish to quantise this system with group averaging for general $p$
and~$q$, using test states built from eigenstates of the harmonic
oscillator Hamiltonians that arise in the oscillator representation of
$\SLtwor$~\cite{Howe}. When $p\ge2$, $q\ge2$ and $p+q \equiv 0
\pmod2$, we obtain a quantum theory in which the classical
$\mathfrak{o}(p,q)$ observables are promoted into a
nontrivially-represented operator algebra. When $(p,q) = (1,3)$ or
$(3,1)$, we obtain a quantum theory with a one-dimensional physical
Hilbert space that is annihilated by all the $\mathfrak{o}(p,q)$
observables. For other values of $p$ and $q$ we recover no physical
Hilbert space. In particular, for $p=q=1$ the group averaging
converges to an \emph{indefinite\/} sesquilinear form, in a sense
strong enough for the uniqueness theorem of \cite{GM2} to imply that
the system admits no rigging maps. This is the first example known to
us in which group averaging fails to produce a Hilbert space owing to
indefiniteness of the would-be inner product.

We show further that all our group averaging quantum theories 
can be obtained within Ashtekar's algebraic
quantisation \cite{Ash1,Ash2}, 
using the $\mathfrak{o}(p,q)$ observables to determine the physical
inner product, 
and we display explicitly the correspondence
between the two schemes. We have not gained sufficient control
over the $\mathfrak{o}(p,q)$ algebra 
to ascertain whether algebraic quantisation 
might for some $p$ and $q$ 
yield also quantum theories not recovered by the group
averaging, but we show that this does not happen for
$p+q \equiv 1 \pmod2$, nor does it happen for 
$p+q \equiv 0 \pmod2$ if $p\le3$ and $q\le3$. 

We also give a detailed description of the classical reduced phase
space. The reduced phase space contains 
a symplectic manifold if and only if $p\ge2$ and $q\ge2$. 
This manifold is 
separated by the $\mathfrak{o}(p,q)$ observables,
and it is connected if and only if $p\ge3$ and $q\ge3$. 
This suggests that 
interesting quantum theories should exist only
when $p\ge2$ and $q\ge2$, possibly with some subtleties when
$\min(p,q) = 2$. As outlined above, this agrees with our findings.

The rest of the paper is as follows. Section \ref{sec:classical}
introduces and analyses the classical system. Section \ref{sec:AQ}
discusses algebraic quantisation, laying out the task for general $p$
and $q$ and completing it for $\max(p,q) \le 3$. Refined algebraic
quantisation with group averaging is carried out in section 
\ref{sec:RAQus} for $\min(p,q) \ge3$ and in section 
\ref{sec:RAQ-low} for other values of $p$ and~$q$. 

Section \ref{sec:discussion} presents a summary and concluding
remarks. Appendix \ref{app:SLtwor} collects some basic properties of
$\SLtwor$, and appendices \ref{app:lin-ind}--\ref{app:convergence}
contain 
the proofs of several technical results stated in the main text.

\section{Classical system}
\label{sec:classical}

In this section we analyse a classical constrained system with the
unreduced phase space $T^*\mathbb{R}^{p+q}$, where $p\ge1$ and
$q\ge1$. The system was introduced for $p=q=2$ in~\cite{MRT}, and our
discussion of the gauge transformations and the distinguished
$\mathfrak{o}(p,q)$ observables generalises the observations of
\cite{MRT} in a straightforward manner. We shall however show that the
structure of the reduced phase space depends sensitively on $p$
and~$q$.

\subsection{The system}
\label{subsec:model}

The system is defined by the action
\be
S=\int 
dt \, 
\bigl( 
\boldsymbol{p} \cdot \dot{\boldsymbol{u}}
+ 
\boldsymbol{\pi} \cdot \dot{\boldsymbol{v}}
-NH_1 -MH_2 -\lambda D
\bigr)
\ \ ,
\label{eq:act1}
\ee
where $\boldsymbol{u}$ and 
$\boldsymbol{p}$ are real vectors of dimension~$p\ge1$,
$\boldsymbol{v}$ and $\boldsymbol{\pi}$ are real vectors of dimension~$q\ge1$,
and the overdot denotes differentiation with respect
to~$t$. $\boldsymbol{p}$~and $\boldsymbol{\pi}$ are respectively the momenta
conjugate to $\boldsymbol{u}$ and~$\boldsymbol{v}$,
the symplectic structure is
\begin{equation}
\Omega = \sum_{i=1}^p dp_i \wedge du_i
+
\sum_{j=1}^q d\pi_j \wedge dv_j
\ \ ,
\label{eq:Omega-on-Gamma}
\end{equation}
and the phase space is
$\Gamma:=T^*\mathbb{R}^{p+q} \simeq \mathbb{R}^{2(p+q)}$.
$N$, $M$ and $\lambda$ are Lagrange multipliers
associated with the constraints
\be
H_1 &:=& \tfrac12 (\boldsymbol{p}^2-\boldsymbol{v}^2)
\ \ ,
\nn
\\
H_2 &:=& \tfrac12 (\boldsymbol{\pi}^2-\boldsymbol{u}^2)
\ \ ,
\nn
\\
D &:=& \boldsymbol{u}\cdot\boldsymbol{p}- \boldsymbol{v}\cdot\boldsymbol{\pi}
\ \ .
\label{eq:const}
\ee
The Poisson algebra of
the constraints is the $\sltwor$ Lie algebra 
(see appendix~\ref{app:SLtwor}), 
\be
\{H_1\, , H_2\}&=&D
\ \ ,
\nn
\\
\{H_1\, , D\}  &=&-2H_1
\ \ ,
\nn
\\
\{H_2\, , D\}  &=&2H_2
\ \ ,
\label{eq:algebra}
\ee
and the system is a first class constrained
system~\cite{Dir3, Hen}. 
The finite gauge transformations on $\Gamma$
generated by the constraints are
\be
\left(
\begin{array}{c}
\boldsymbol{u}\\
\boldsymbol{p}\\
\end{array}
\right)\mapsto  g\left(
\begin{array}{c}
\boldsymbol{u}\\
\boldsymbol{p}\\
\end{array}
\right), \hspace{5ex}
\left(
\begin{array}{c}
\boldsymbol{\pi}\\
\boldsymbol{v}\\
\end{array}
\right)\mapsto  g\left(
\begin{array}{c}
\boldsymbol{\pi}\\
\boldsymbol{v}\\
\end{array}
\right),
\label{eq:gauge-transf}
\ee
where $g$ is an 
$\SLtwor$ matrix. The gauge group is thus $\SLtwor$. As the
Hamiltonian is a sum of the constraints, the constraints entirely
determine the dynamics.

\subsection{Classical observables}
\label{subsec:class-obs}

Recall that an observable is a function on $\Gamma$ whose
Poisson brackets with the first class
constraints vanish when the first class constraints hold~\cite{Hen}.
Consider on
$\Gamma$ the functions $\mathcal{O}_{kj}:= x_k \times x_j$, where
$x_k = (u_k,p_k)^T$ for $1\le k \le p$,
$x_{p+k} = (\pi_k,v_k)^T$ for $1\le k \le q$, and
the cross stands for the scalar-valued cross product
on~$\BbbR^2$.
As the $\SLtwor$ action on $\BbbR^2$
preserves areas,
(\ref{eq:gauge-transf}) shows that $\mathcal{O}_{kj}$ are
invariant under the gauge transformations.
Hence $\mathcal{O}_{kj}$
are observables.
The Poisson algebra of these observables is
the $\mathfrak{o}(p,q)$ Lie algebra, 
\be
\{\mathcal{O}_{ij}} , {\mathcal{O}_{kl}\}
=g_{ik}\mathcal{O}_{jl}-
g_{il}\mathcal{O}_{jk}+g_{jl}\mathcal{O}_{ik}
- g_{jk}\mathcal{O}_{il}
\ \ ,
\label{eq:sopq-class}
\ee
where
\be
g_{ik}
=
\mathrm{diag}(\underbrace{1,\ldots,1}_p,\underbrace{-1,\ldots,-1}_q)_{ik}
\ \ .
\ee
The algebra generated by $\{ \mathcal{O}_{ij} \}$ is denoted
by~$\Aclass$. The finite transformations that $\Aclass$ generates on
$\Gamma$ are
\be
\left(
\begin{array}{c}
\boldsymbol{u}\\
\boldsymbol{\pi}\\
\end{array}
\right)\mapsto  R
\left(
\begin{array}{c}
\boldsymbol{u}\\
\boldsymbol{\pi}\\
\end{array}
\right), \hspace{5ex}
\left(
\begin{array}{c}
\boldsymbol{p}\\
\boldsymbol{v}\\
\end{array}
\right)\mapsto  R
\left(
\begin{array}{c}
\boldsymbol{p}\\
\boldsymbol{v}\\
\end{array}
\right),
\label{eq:alg-action}
\ee
where $R$ is an $\mathrm{O}(p,q)$ matrix, 
in the connected component 
$\mathrm{O}_{\mathrm{c}}(p,q)$.
Note that as none of the above relies on the constraints being
satisfied,
the
$\SLtwor$ action (\ref{eq:gauge-transf})
and the
$\mathrm{O}_{\mathrm{c}}(p,q)$-action (\ref{eq:alg-action})
commute on all of~$\Gamma$, not just on the subset where the
constraints hold.

It will be useful to decompose the basis $\{ \mathcal{O}_{ij} \}$ of
$\Aclass$ as $\mathfrak{o}(p,q) =
\mathfrak{o}(p)
\oplus
\mathfrak{o}(q)
\oplus
\mathfrak{p}$,
where $\mathfrak{p}$ is spanned by the observables transverse to
those in the Lie algebra of the maximal compact subgroup
$\mathrm{O}_{\mathrm{c}}(p)\times
\mathrm{O}_{\mathrm{c}}(q)$~\cite{Gil}.
Explicitly, we write
\begin{equation}
\begin{array}{lcll}
A_{ij}
& := &
\mathcal{O}_{ij}
= u_i p_j-u_j p_i
\ ,
\hspace{7ex}
&
1\le i \le p,
\ \
1\le j\le p
\ ;
\\
B_{ij}
& := &
\mathcal{O}_{p+i,p+j}
= v_i \pi_j-v_j \pi_i
\ ,
&
1\le i \le q,
\ \
1 \le j\le q
\ ;
\\
C_{ij}
& := &
\mathcal{O}_{i,p+j}
= u_i v_j-p_i \pi_j
\ ,
&
1\leq i\leq p,
\ \
1\leq j\leq q
\ ,
\end{array}
\label{eq:obser}
\end{equation}
where
$A_{ij} \in \mathfrak{o}(p)$,
$B_{ij} \in \mathfrak{o}(q)$
and
$C_{ij} \in \mathfrak{p}$.

Other observables of interest in $\Aclass$ are the 
Casimir elements of the universal enveloping algebra of
$\mathfrak{o}(p,q)$~\cite{AMMP}. We consider only the
quadratic Casimir observable, 
\begin{eqnarray}
\mathcal{C}
&:=&
\frac12
\sum_{ijkl} g_{ij}g_{kl} \mathcal{O}_{ik}
\mathcal{O}_{jl}
\nonumber
\\
&=&
\sum_{i < j} {(A_{ij})^2}
\
+
\sum_{i < j} {(B_{ij})^2}
\
-
\sum_{i,j} {(C_{ij})^2}
\nonumber
\\
& = &
- 4H_1H_2 - D^2
\ \ ,
\label{eq:class-casimir}
\end{eqnarray}
where the last equality follows by direct computation.
When the constraints hold, $\mathcal{C}$ thus vanishes.

\subsection{Reduced phase space}
\label{subsec:red-phasespace}

Let $\barGamma$ be the subset of $\Gamma$ where the constraints
hold. The reduced phase space, denoted by~$\Mred$, is the
quotient of $\barGamma$ under the gauge action~(\ref{eq:gauge-transf}).
As
the Hamiltonian is a linear combination of the constraints, there is no
dynamics on~$\Mred$, and $\Mred$ can be identified with the space of
classical solutions. As the functions in $\Aclass$ are gauge invariant,
they project to functions on~$\Mred$: We use for these functions the same
symbols.

For $p=q=2$, the generic sectors of $\Mred$ were found in
\cite{MRT,TruSL} and the global properties of $\Mred$ were exhibited
in~\cite{LouRov}. We now analyse $\Mred$ for general $p\ge1$ and
$q\ge1$.


$\barGamma$ is clearly connected.
Hence also $\Mred$ is connected.

To proceed, we decompose $\barGamma$ into three subsets. Let
$\barGammanought = \{ \qnought \}$, where $\qnought$ is the origin
of~$\Gamma$, 
$\boldsymbol{u} = \boldsymbol{p} = \boldsymbol{0} = \boldsymbol{v} =
\boldsymbol{\pi}$. Let $\barGammaex$ contain all other points of
$\barGamma$ at which at least one of the pairs 
$(\boldsymbol{u},\boldsymbol{p})$ and
$(\boldsymbol{v},\boldsymbol{\pi})$ is linearly dependent. Finally, let
$\barGammareg$ contain the rest of~$\barGamma$.
We refer to $\barGammaex$ and $\barGammareg$ as respectively the
``exceptional'' and ``regular'' parts of~$\barGamma$.
We show in appendix \ref{app:lin-ind}
that the gradients of the constraints
are all vanishing on~$\barGammanought$,
linearly
dependent but not all vanishing on~$\barGammaex$, and linearly
independent on~$\barGammareg$.
$\barGammanought$~and $\barGammaex$ are nonempty for all $p$ and~$q$,
while $\barGammareg$ is nonempty if and only if $p\ge2$ and $q\ge2$.

As $\barGammanought$, $\barGammaex$ and $\barGammareg$ are
preserved by the gauge transformations, they project onto disjoint
subsets of~$\Mred$. We denote these sets respectively by
$\Mnought$, $\Mex$ and $\Mreg$ and analyse each in turn.

\subsubsection{$\Mnought$}

$\Mnought$ contains only one point, the projection of~$\qnought$.
All observables in $\Aclass$ vanish on~$\Mnought$.

\subsubsection{$\Mex$}

As $\qnought\notin\barGammaex$, the constraints
$H_1=0=H_2$ show that all points in $\barGammaex$ have
$(\boldsymbol{u},\boldsymbol{p}) \ne (\boldsymbol{0},\boldsymbol{0})$
and
$(\boldsymbol{v},\boldsymbol{\pi}) \ne (\boldsymbol{0},\boldsymbol{0})$.
Given a point at which
the pair $(\boldsymbol{u},\boldsymbol{p})$ is linearly
dependent, there thus exists a gauge-equivalent point with
$\boldsymbol{u}=\boldsymbol{0}$ and $\boldsymbol{p}^2=1$,
at which the constraints imply
$\boldsymbol{\pi}=\boldsymbol{0}$ and $\boldsymbol{v}^2=1$.
Given a point at which
the pair $(\boldsymbol{v},\boldsymbol{\pi})$
is linearly dependent,
a similar argument shows that there exists a
gauge-equivalent point at which
$\boldsymbol{\pi}=\boldsymbol{0}$,
$\boldsymbol{v}^2=1$,
$\boldsymbol{u}=\boldsymbol{0}$ and
$\boldsymbol{p}^2=1$.
Thus, each point in $\barGammaex$ is
gauge-equivalent to a point that satisfies
\begin{equation}
\boldsymbol{v}^2 = \boldsymbol{p}^2 = 1
\ \ ,
\ \
\boldsymbol{u}= \boldsymbol{0} = \boldsymbol{\pi}
\ \ .
\label{eq:Mex-char}
\end{equation}
It follows that both the pair
$(\boldsymbol{u},\boldsymbol{p})$ and the pair $(\boldsymbol{v},\boldsymbol{\pi})$ are
linearly dependent on~$\barGamma$.

The gauge transformations that preserve the set (\ref{eq:Mex-char})
act on it either trivially or by
\begin{equation}
(\boldsymbol{v},\boldsymbol{p}) \mapsto (-\boldsymbol{v},-\boldsymbol{p})
\ \ .
\label{eq:Mex-zedtwo}
\end{equation}
$\Mex$ can therefore be represented as the
quotient of the set~(\ref{eq:Mex-char}),
with topology $S^{p-1} \times S^{q-1}$, under the
$\BbbZ_2$ action generated
by~(\ref{eq:Mex-zedtwo}). If in particular $p=1$ (respectively
$q=1$), $\Mex$~has topology $S^{q-1}$ ($S^{p-1}$).
If $p=q=1$, $\Mex$~contains just two points.

Other representations of $\Mex$ are obtained by replacing in
(\ref{eq:Mex-char}) the first equations by 
$\boldsymbol{v}^2 = \boldsymbol{p}^2 = r$,
where $r$ is an arbitrary prescribed positive number. This shows that
in the topology of $\Mred$ induced from~$\Gamma$, 
every open set that includes $\Mnought$ includes also~$\Mex$. 

Equations (\ref{eq:obser}) and (\ref{eq:Mex-char})
show that all observables in
$\Aclass$ vanish on~$\Mex$.
Equations (\ref{eq:Omega-on-Gamma}) and (\ref{eq:Mex-char}) show that the
projection of the symplectic form $\Omega$ vanishes on~$\Mex$. We
refer to $\Mex$ as the ``exceptional'' part of~$\Mred$.

\subsubsection{$\Mreg$}
\label{subsubsec:Mreg}

When $p=1$ or $q=1$ (or both), $\barGammareg$ and hence also $\Mreg$
are empty. We now assume $p\ge2$ and $q\ge2$.

We show in appendix \ref{app:lin-ind} that 
the gradients of the
constraints are linearly independent
on~$\barGammareg$. It follows (\cite{Hen},
Section 1.1.2 and Appendix~2A) that $\Mreg$ is a manifold of dimension
$2p+2q - 6$ with a symplectic form induced from~$\Gamma$. 
We refer to $\Mreg$ as the ``regular'' part of~$\Mred$.

Given a point in~$\barGammareg$, the linear independence of the
pair $(\boldsymbol{u},\boldsymbol{p})$ implies that there
exists a gauge-equivalent point
at which $\boldsymbol{u} \cdot \boldsymbol{p} =0$ and $\boldsymbol{u}^2 = \boldsymbol{p}^2>0$.
The constraints imply that at this point
$\boldsymbol{v} \cdot \boldsymbol{\pi} =0$,
$\boldsymbol{v}^2 = \boldsymbol{p}^2$ and
$\boldsymbol{\pi}^2 = \boldsymbol{u}^2$.
Hence each point in $\barGammareg$ is
gauge-equivalent to a point that satisfies
\begin{equation}
\boldsymbol{u}^2 = \boldsymbol{p}^2 = \boldsymbol{v}^2 = \boldsymbol{\pi}^2
>0
\ \ ,
\ \
\boldsymbol{u} \cdot \boldsymbol{p} = \boldsymbol{v} \cdot \boldsymbol{\pi}
= 0
\ \ .
\label{eq:Mreg-char}
\end{equation}
The gauge transformations that preserve the set
(\ref{eq:Mreg-char}) are (\ref{eq:gauge-transf}) with
\begin{equation}
g=
\left(\begin{array}{cc}
\cos\theta  & \sin\theta
\\
-\sin\theta & \cos\theta
\end{array}\right)
\ \ ,
\label{eq:Mreg-uone}
\end{equation}
where $0\le\theta < 2\pi$.
It follows that $\Mreg$ can be represented as the quotient of the set
(\ref{eq:Mreg-char}) under the
$U(1)$ action given by (\ref{eq:gauge-transf}) and~(\ref{eq:Mreg-uone}).

We show in appendix \ref{app:separation} that $\Aclass$
separates~$\Mreg$: Given two distinct points in~$\Mreg$, there exist
functions in $\Aclass$ that take distinct values at the two points.

For $p=q=2$, $\Mreg$
consists of four connected components~\cite{LouRov,MRT,TruSL}, 
which can be pairwise joined
into two connected symplectic manifolds by adding certain points
from $\Mex$~\cite{LouRov}. 


Suppose $p=2$ and $q>2$. Within each gauge equivalence class
in~(\ref{eq:Mreg-char}), there is a unique representative at which
$p_1=0$ and
$u_1>0$. It follows that at this point $p_2\ne0$ and $u_2=0$.
A gauge transformation by $g =
\mathrm{diag}(|p_2|, |p_2|^{-1})$ brings this point to
\begin{equation}
\boldsymbol{v}^2=1
\ \ , \ \
\boldsymbol{v} \cdot \boldsymbol{\pi} = 0
\ \ , \ \
\boldsymbol{\pi}^2 >0
\ \ , \ \
\boldsymbol{p} = (0,\epsilon)
\ \ , \ \
\boldsymbol{u} = (|\boldsymbol{\pi}|, 0)
\ \ ,
\label{eq:Mreg-char-23}
\end{equation}
where $\epsilon = \pm1$. For each~$\epsilon$,
the set
(\ref{eq:Mreg-char-23}) is recognised as the cotangent bundle
over~$S^{q-1}$, with the zero fibres omitted. Hence $\Mreg$
consists of two connected components, given by
(\ref{eq:Mreg-char-23}) with the respective values
of~$\epsilon$. 
Equations (\ref{eq:Omega-on-Gamma}) and (\ref{eq:Mreg-char-23})
show that
the symplectic structure of
this cotangent bundle description is precisely the symplectic
structure induced from~$\Gamma$.
For each~$\epsilon$, it is possible to include
the zero fibres by allowing $\boldsymbol{\pi}^2=0$ in
(\ref{eq:Mreg-char-23}); this means adding from $\Mex$ the subset
represented uniquely by (\ref{eq:Mex-char}) with $\boldsymbol{p} =
(0,\epsilon)$ and
$\boldsymbol{\pi} = \boldsymbol{0}$. 
Note that because of the identification
(\ref{eq:Mex-zedtwo}) in~(\ref{eq:Mex-char}), this subset of
$\Mex$ is the same for both
signs of~$\epsilon$. The mechanism of pairwise smoothly 
joining the disconnected sectors for $q=2$ \cite{LouRov} is not
available now because the fibres without origin are disconnected for $q=2$
but connected for $q>2$.

The case $q=2$ and $p>2$ is isomorphic to $p=2$ and $q>2$.

When $p>2$ and $q>2$, $\Mreg$ is connected. We have not found a
simpler description of the global properties in this case. Convenient
local gauge fixings are introduced in
appendix~\ref{app:separation}.

\section{Algebraic quantisation}
\label{sec:AQ}

In this section we apply the algebraic quantisation framework 
of~\cite{Ash1}, 
adopting $\Aclass$ as the classical
observable algebra whose complex conjugation relations are promoted
into adjointness relations. 
Seeking solutions to the quantum constraints by
separation of variables, we show in subsection \ref{subsec:alg-gen-pq}
that necessary conditions for obtaining a quantum theory with a
nontrivially-represented observable algebra are
$p\ge2$, $q\ge2$ and $p + q \equiv 0 \pmod2$.
The case $p=q=2$ was analysed in
\cite{LouRov,MRT}. In subsection \ref{subsec:pq3}
we complete the quantisation for
$p=q=3$. 

Detailed expositions of algebraic quantisation can be found
in~\cite{Ash1,Ash2}.

\subsection{Setup for $p\ge1$ and $q\ge1$}
\label{subsec:alg-gen-pq}

We take the
elementary
`position' and `momentum' operators to act on smooth functions
$\Psi(\boldsymbol{u},\boldsymbol{v})$ as 
\be
\hat{\boldsymbol{u}}\Psi(\boldsymbol{u},\boldsymbol{v})
=
\boldsymbol{u}\Psi(\boldsymbol{u},\boldsymbol{v}),
\hspace{5ex}
\hat{\boldsymbol{p}} \, \Psi(\boldsymbol{u},\boldsymbol{v})
=
-i\boldsymbol{\nabla}_u \Psi(\boldsymbol{u},\boldsymbol{v}),
\nn
\\
\hat{\boldsymbol{v}} \, \Psi(\boldsymbol{u},\boldsymbol{v})
=
\boldsymbol{v}\Psi(\boldsymbol{u},\boldsymbol{v}),
\hspace{5ex}
\hat{\boldsymbol{\pi}} \Psi(\boldsymbol{u},\boldsymbol{v})
=
-i\boldsymbol{\nabla}_v \Psi(\boldsymbol{u},\boldsymbol{v}),
\label{eq:operator}
\ee
so that
$[\hat{u}_k, \hat{p}_j] = i \delta_{kj}$
and
$[\hat{v}_k, \hat{\pi}_j] = i \delta_{kj}$.
Inserting these operators
into the classical 
constraints (\ref{eq:const}) and making a judicious ordering
choice, we obtain the quantum constraints
\begin{subequations}
\label{eq:con-op}
\be
\hat{H}_1& := & -\tfrac12 \! 
\left( \Delta_u+\boldsymbol{v}^2 \right)
\ \ ,
\\
\hat{H}_2 & := & -\tfrac12 \! 
\left( \Delta_v+\boldsymbol{u}^2 \right)
\ \ ,
\\
\hat{D} & := &
-i\left(\boldsymbol{u}\cdot\boldsymbol{\nabla}_u
-\boldsymbol{v}\cdot\boldsymbol{\nabla}_v
+
\frac{p-q}{2}
\right)
\ \ ,
\ee
\end{subequations}
where $\Delta_u$ (respectively $\Delta_v$) stands for the Laplacian in
$\boldsymbol{u}$~($\boldsymbol{v}$). The non-derivative term in $\hat{D}$
is needed to make the
commutators close as the $\sltwor$ Lie algebra,
\be
\com{\hat{H}_1}{\hat{H}_2} & = & i\hat{D}
\ \ ,
\nn
\\
\com{\hat{H}_1}{\hat{D}} & = & -2i\hat{H}_1
\ \ ,
\nn
\\
\com{\hat{H}_2}{\hat{D}} & = & +2i\hat{H}_2
\ \ .
\ee

We define the quantum observables $\hat{\mathcal{O}}_{ij}$ by
substituting the elementary quantum operators
(\ref{eq:operator}) in the
expressions of the
classical observables~$\mathcal{O}_{ij}$. These quantum observables
commute with the quantum constraints~(\ref{eq:con-op}), and their
commutators form the 
$\mathfrak{o}(p,q)$ Lie algebra, obtained by hatting
(\ref{eq:sopq-class}) and multiplying the right-hand side by~$i$. As
$\mathcal{O}_{ij}$ are real, we introduce on the algebra generated by 
$\{ \hat{\mathcal{O}}_{ij} \}$
an antilinear involution
by $\hat{\mathcal{O}}^{\star}_{ij} =
\hat{\mathcal{O}}_{ij}$. We denote the resulting star-algebra of quantum
observables by~$\Aphystar$. 

Following~(\ref{eq:obser}), we decompose the basis of
$\Aphystar$ as
\begin{equation}
\begin{array}{lcll}
\hat{A}_{ij}
& := &
\hat{\mathcal{O}}_{ij}
= -i \left( u_i \partial_{u_j} - u_j \partial_{u_i} \right)
\ ,
\hspace{7ex}
&
1\le i \le p,
\ \
1\le j \le p
\ ;
\\
\hat{B}_{ij}
& := &
\hat{\mathcal{O}}_{p+i,p+j}
=
-i \left( v_i \partial_{v_j} - v_j  \partial_{v_i} \right)
\ ,
&
1\le i \le q,
\ \
1 \le j \le q
\ ;
\\
\hat{C}_{ij}
& := &
\hat{\mathcal{O}}_{i,p+j}
= u_i v_j + \partial_{u_i} \partial_{v_j}
\ ,
&
1\leq i\leq p,
\ \
1\leq j\leq q
\ .
\end{array}
\label{eq:q-obser}
\end{equation}
The quantum quadratic Casimir observable is
\begin{eqnarray}
\hat{\mathcal{C}}
&:=&
\frac12
\sum_{ijkl} g_{ij}g_{kl}
\hat{\mathcal{O}}_{ik}
\hat{\mathcal{O}}_{jl}
\nonumber
\\
&=&
\sum_{i < j} {(\hat{A}_{ij})^2}
\
+
\sum_{i < j}{(\hat{B}_{ij})^2}
\
-
\sum_{i,j} {(\hat{C}_{ij})^2}
\nonumber
\\
& = &
- 2 \bigl( \hat{H}_1\hat{H}_2 + \hat{H}_2\hat{H}_1 \bigr)
- \hat{D}^2
- \textstyle{\frac14} (p+q)(p+q-4)
\ \ ,
\label{eq:pq33-quantum-casimir}
\end{eqnarray}
where the last equality follows by direct computation. 
In contrast to the classical Casimir~(\ref{eq:class-casimir}), 
$\hat{\mathcal{C}}$ vanishes on states annihilated by the
constraints only for $p+q=4$. 

We seek states annihilated by the constraints,
\begin{equation}
\hat{H}_1 \Psi(\boldsymbol{u},\boldsymbol{v})=0
\ \ ,
\ \
\hat{H}_2 \Psi(\boldsymbol{u},\boldsymbol{v})=0
\ \ ,
\ \
\hat{D} \Psi(\boldsymbol{u},\boldsymbol{v})=0
\ \ ,
\ \
\label{eq:qconstraints}
\end{equation}
by separation of variables. If $p \ge 2$ and $q\ge2$, we make the
ansatz
\begin{equation}
\Psi(\boldsymbol{u},\boldsymbol{v}) =\psi(u,v)
Y_{lk_u}
\bigl({\theta}^{(u)}\bigr)
Y_{jk_v}
\bigl({\theta}^{(v)}\bigr)
\ \ ,
\label{eq:separation-ansatz}
\end{equation}
where $u:=|\boldsymbol{u}|$,
$v:=|\boldsymbol{v}|$ and $Y_{lk_u}\bigl({\theta}^{(u)}\bigr)$
(respectively $Y_{jk_v}\bigl({\theta}^{(v)}\bigr)$)
are the
spherical harmonics on unit $S^{p-1}$ in $\boldsymbol{u}$
($S^{q-1}$ in $\boldsymbol{v}$)~\cite{Bate,Vil}.
Here ${\theta}^{(u)}$ denotes the coordinates
on~$S^{p-1}$, the index $l$ ranges over non-negative integers, the
eigenvalue of the scalar Laplacian on $S^{p-1}$ is $-l(l+p-2)$, the
index $k_u$ labels the degeneracy for each~$l$, and
similarly for the quantities appearing in
$Y_{jk_v}\bigl({\theta}^{(v)}\bigr)$.
We extend the ansatz (\ref{eq:separation-ansatz}) to
$p=1$, in which case
${\theta}^{(u)} := u_1 / u \in
\{1,-1\}$,
$l \in \{0,1\}$,
the index $k_u$ takes only a single value and can be dropped,
and the spherical
harmonics are $Y_l \bigl({\theta}^{(u)}\bigr) :=
\bigl({\theta}^{(u)}\bigr)^l / \sqrt{2}$, and similarly for
$q=1$.
For all $p\ge1$ and $q\ge1$,
equations (\ref{eq:qconstraints}) then reduce to
\begin{subequations}
\be
\left[
\frac{1}{u^{p-3}}
\frac{\partial}{\partial u}
\left(u^{p-1}\frac{\partial}{\partial u}\right)
-l(l+p-2)+u^2v^2
\right]
\psi(u,v)
& = &
0
\ \ ,
\label{eq:polarop-q1}
\\
\left[
\frac{1}{v^{q-3}}
\frac{\partial}{\partial v}
\left(v^{q-1}\frac{\partial}{\partial v} \right)
-j(j+q-2)+u^2v^2
\right]
\psi(u,v)
& = &
0
\ \ ,
\label{eq:polarop-q2}
\\
\left(
u\frac{\partial}{\partial u}
- v\frac{\partial}{\partial v}
+ \frac{p-q}{2}
\right)
\psi(u,v)
& = &
0
\ \ .
\label{eq:polarop-linear}
\ee
\end{subequations}

The general solution to (\ref{eq:polarop-linear}) is 
$\psi(u,v)=u^{(2-p)/2}v^{(2-q)/2}\chi(\zeta)$,
where $\zeta := uv$. 
Substituting this in (\ref{eq:polarop-q1}) and (\ref{eq:polarop-q2}),
we find that the indices satisfy
\begin{equation}
2l+p = 2j+q
\label{eq:index-equality}
\end{equation}
and $\chi(\zeta)$ satisfies the Bessel equation
of order $l + (p-2)/2$~\cite{Bate}. 

Equation (\ref{eq:index-equality}) shows that solutions exist 
only when $p + q \equiv 0 \pmod2$. If $p=1$ or $q=1$, inspection
of (\ref{eq:index-equality}) further shows that solutions exist 
only when $(p,q) = (1,1)$, $(1,3)$ or $(3,1)$. 
Let us consider these exceptional
cases first.

When $p=q=1$, the linearly independent solutions are
$\Psi_\pm := \exp(\pm i u_1 v_1)$,
$\Aphystar$ is generated by
the single observable $\hat{C}_{11}$, and
$\hat{C}_{11} \Psi_\pm = \pm
i \Psi_\pm$. The representation of
$\Aphystar$ on $\Vphys := \mathrm{span}
\{ \Psi_\pm \}$ is irreducible, but the only sesquilinear forms in which 
$\hat{C}_{11}$ is symmetric have indefinite signature. 

When $p=1$ and $q=3$, the only (smooth) solution is
$\Psi_0 :=  v^{-1}\sin(u_1v)$, which is annihilated by all operators
in~$\Aphystar$. Promoting $\mathrm{span} \{ \Psi_0 \}$ into a
one-dimensional Hilbert space gives thus a quantum theory in which
$\Aphystar$ is represented trivially. The situation for $p=3$ and $q=1$ is
similar.

We therefore see that necessary conditions for obtaining
a quantum theory with a nontrivial representation
of $\Aphystar$
are $p\ge2$, $q\ge2$ and $p + q \equiv 0 
\pmod2$. When these conditions hold, we have found for the quantum
constraints the linearly independent solutions
\begin{equation}
\Psi_{ljk_uk_v}
:=
\delta_{2l+p,2j+q}
\,
u^{(2-p)/2}v^{(2-q)/2}
J_{l + (p-2)/2}(uv)
Y_{lk_u}
\bigl({\theta}^{(u)}\bigr)
Y_{jk_v}
\bigl({\theta}^{(v)}\bigr)
\ \ ,
\label{eq:physstates}
\end{equation}
where
$J_{l + (p-2)/2}$ is the
Bessel function of the first kind~\cite{Bate}. The
Bessel function of the second kind has been excluded to make
$\Psi_{ljk_uk_v}$ smooth at $uv=0$. The motivation for this exclusion
may be debatable within algebraic quantisation, 
but we shall see that it is precisely
the smooth solutions (\ref{eq:physstates}) that will
emerge from group averaging in sections \ref{sec:RAQus}
and~\ref{sec:RAQ-low}. 

To proceed, we would need to examine the representation of
$\Aphystar$ on $\mathrm{span} \{\Psi_{ljk_uk_v}\}$. The representation
of the $\mathfrak{o}(p) \oplus \mathfrak{o}(q)$ subalgebra is given
directly by its representation on the spherical
harmonics~\cite{Bate,Vil}, but the observables $\hat{C}_{ij}$ mix the
states in a more complicated way. The special case $p=q=2$ was
analysed in~\cite{LouRov,MRT}. In subsection \ref{subsec:pq3} we
address the special case $p=q=3$.

\subsection{Completion for $p=q=3$}
\label{subsec:pq3}

When $p=q=3$, the states (\ref{eq:physstates}) can be written as
\begin{equation}
\Psi_{lmn}=j_l(uv)
Y_{lm} 
\bigl({\theta}^{(u)}\bigr)
Y_{ln}
\bigl({\theta}^{(v)}\bigr)
\ \ ,
\label{eq:phys3-3}
\end{equation}
where $l$ ranges over nonnegative integers,
$j_l(uv)$ is the spherical Bessel function of the first kind
of order $l$~\cite{Bate},
$m$ and $n$ are integers satisfying
$|m| \le l$ and $|n| \le l$
and the $Y$'s are the
usual spherical harmonics on $S^2$~\cite{Bate}. We write
$\Vphys := \mathrm{span} \{ \Psi_{lmn} \}$.

We introduce for $\Aphystar$ the basis
\be
\hat{L}_3         & := & \hat{A}_{12} \ \ ,\nn \\
\hat{L}_\pm       & := & \hat{A}_{23}\pm i\hat{A}_{31} \ \ , \nn  \\
\hat{J}_3         & := & \hat{B}_{12} \ \ ,\nn\\
\hat{J}_\pm       & := & \hat{B}_{23}\pm i\hat{B}_{31} \ \ , \nn   \\
\hat{C}_{0}       & := & \hat{C}_{33} \ \ ,\nn\\
\hat{C}_{1}^{\pm} & := & \hat{C}_{31}\pm i\hat{C}_{32} \ \ , \nn \\
\hat{C}_{2}^{\pm} & := & \hat{C}_{13}\pm i\hat{C}_{23} \ \ , \nn \\
\hat{C}_{3}^{\pm} & := &
(\hat{C}_{11}+\hat{C}_{22})\pm i(\hat{C}_{21}-\hat{C}_{12}) \ \ , \nn \\
\hat{C}_{4}^{\pm} & := &
(\hat{C}_{11}-\hat{C}_{22})\pm i(\hat{C}_{21}+\hat{C}_{12}) \ \ .
\label{eq:comb}
\ee
Note that the $\hat{L}$'s (respectively $\hat{J}$'s) are a standard
raising and lowering operator
basis for the $\mathfrak{o}(3)$ algebra in
$\boldsymbol{u}$ ($\boldsymbol{v}$)~\cite{Bate}. 
The action of the basis 
(\ref{eq:comb}) on 
$\Vphys$ can be computed from standard properties
of the spherical harmonics and spherical Bessel functions
\cite{Bate,arfken} 
and is displayed in Table~\ref{table:observables}.
It follows that $\Vphys$ is invariant under~$\Aphystar$. 
We show in appendix
\ref{app:pq33-algebra} that the representation of $\Aphystar$ on
$\Vphys$ is irreducible.

\begin{table}[p]
\be
\hat{L}_{3}\Psi_{lmn} & = & m \Psi_{lmn}                                 \nn \\
\hat{J}_{3}\Psi_{lmn} & = & n \Psi_{lmn}                                 \nn \\
\hat{L}_{\pm}\Psi_{lmn} & = & \sqrt{(l\pm m+1)(l\mp m)}\Psi_{l,m\pm 1,n}  \nn \\
\hat{J}_{\pm}\Psi_{lmn} & = & \sqrt{(l\pm n+1)(l\mp n)}\Psi_{l,m,n\pm 1}  \nn \\
\hat{C}_{0}\Psi_{lmn} & = & \frac{\sqrt{(l-m+1)(l+m+1)(l-n+1)(l+n+1)}}{2l+3}\Psi_{l+1,m,n}\nn\\
                      &  & + \frac{\sqrt{(l-m)(l+m)(l-n)(l+n)}}{2l-1}\Psi_{l-1,m,n}                    \nn \\
\hat{C}_{1}^{\pm}\Psi_{lmn} & = & \mp \frac{\sqrt{(l-m+1)(l+m+1)(l\pm n+1)(l\pm n+2)}}{2l+3}\Psi_{l+1,m,n\pm 1} \nn\\
                            & \ & \pm  \frac{\sqrt{(l-m)(l+m)(l\mp n)(l\mp n-1)}}{2l-1}\Psi_{l-1,m,n\pm 1}   \nn\\
\hat{C}_{2}^{\pm}\Psi_{lmn} & = & \mp \frac{\sqrt{(l\pm m+1)(l\pm m+2)(l-n+1)(l+n+1)}}{2l+3}\Psi_{l+1,m\pm 1,n} \nn\\
                            & \ & \pm  \frac{\sqrt{(l\mp m)(l\mp m-1)(l-n)(l+n)}}{2l-1}\Psi_{l-1,m\pm 1,n}                   \nn \\
\hat{C}_{3}^{\pm}\Psi_{lmn} & = & - \frac{\sqrt{(l\pm m+1)(l\pm m+2)(l\mp n+1)(l\mp n+2)}}{2l+3}\Psi_{l+1,m\pm 1,n\mp 1} \nn\\
                            & \ & -  \frac{\sqrt{(l\mp m)(l\mp m-1)(l\pm n)(l\pm n-1)}}{2l-1}\Psi_{l-1,m\pm 1,n\mp 1}                    \nn \\
\hat{C}_{4}^{\pm}\Psi_{lmn} & = & + \frac{\sqrt{(l\pm m+1)(l\pm m+2)(l\pm n+1)(l\pm n+2)}}{2l+3}\Psi_{l+1,m\pm 1,n\pm 1} \nn\\
                            & \ & +  \frac{\sqrt{(l\mp m)(l\mp m-1)(l\mp n)(l\mp n-1)}}{2l-1}\Psi_{l-1,m\pm 1,n\pm 1}  \nn
\ee
\caption{The action of $\Aphystar$ on~$\Vphys$. 
Whenever the indices
of a $\Psi$ on the right-hand side go outside the allowed range,
the numerical coefficient vanishes and the term is
understood as zero.}
\label{table:observables}
\end{table}

The star-relations of the basis (\ref{eq:comb}) read
\be
(\hat{L}_3)^\star         & = & \hat{L}_3 \ \ ,\nn \\
(\hat{L}_\pm)^\star      & = & \hat{L}_\mp \ \ , \nn  \\
(\hat{J}_3)^\star         & = & \hat{J}_3 \ \ ,\nn\\
(\hat{J}_\pm)^\star       & = & \hat{J}_\mp \ \ , \nn   \\
(\hat{C}_{0})^\star      & = & \hat{C}_{0} \ \ ,\nn\\
(\hat{C}_{k}^{\pm})^\star & = & \hat{C}_{k}^{\mp}
\ \ , \ \ 1\le k \le
4 \ \ .
\label{eq:pq33-star}
\ee
{}From Table \ref{table:observables} it follows by
direct computation that these star-relations coincide with the adjoint
relations in the inner product
\begin{equation}
(\Psi_{lmn}\ , \Psi_{l'm'n'})_{\mathrm{AQ}}
:=
(2l+1)\delta_{ll'}\delta_{mm'}\delta_{nn'}
\ \ .
\label{eq:pq33-ip}
\end{equation}
We show in appendix \ref{app:pq33-algebra} that the only inner
products on $\Vphys$ with this property are multiples
of~(\ref{eq:pq33-ip}).

The physical Hilbert space is the Cauchy completion of $\Vphys$ in the
inner product~(\ref{eq:pq33-ip}). It carries by construction a
densely-defined 
representation of $\Aphystar$ in which the quadratic
$\mathfrak{o}(3,3)$ Casimir (\ref{eq:pq33-quantum-casimir}) has the
value~$-3$.

\section{Refined algebraic quantisation for $p\ge3$, $q\ge3$}
\label{sec:RAQus}

We now turn to refined algebraic quantisation. In this section we take
$p\ge3$ and $q\ge3$. The
remaining values of $p$ and $q$ will be treated 
in section~\ref{sec:RAQ-low}.

We employ refined algebraic quantisation with group averaging 
as formulated
in~\cite{GM2}. A~review can be found in \cite{Marolf-MG} and an
outline adapted to the present situation in~\cite{LouRov}.

\subsection{Auxiliary Hilbert space and 
representation of the gauge group}
\label{subsec:auxhil}

We introduce the auxiliary Hilbert space $\Haux \simeq
L^2(\BbbR^{p+q})$ of square integrable functions $\Psi
(\boldsymbol{u},\boldsymbol{v})$ in the inner product
\begin{equation}
(\Psi_1, \Psi_2)_\mathrm{aux}
:=
\int d^p\boldsymbol{u} \, d^q\boldsymbol{v}
\,
\overline{\Psi_1} \Psi_2
\ \ , 
\end{equation}
where the overline denotes complex conjugation. 
The quantum constraints (\ref{eq:con-op}) are essentially self-adjoint
on~$\Haux$, and exponentiating  $-i$ times their algebra yields
a unitary
representation of the universal covering group of $\SLtwor$. Denoting
this representation by~$U$, the group elements
in the Iwasawa decomposition 
(\ref{eq:iwasawa-decomp1}) are
represented by
\begin{subequations}
\label{eq:Iwasawa}
\be
U\bigl(\exp(\mu e^-)\bigr)
& = &
\exp(-i\mu\hat{H}_2)
\ \ ,
\label{eq:Iwasawa1}
\\
U\bigl(\exp(\lambda h)\bigr)
& = &
\exp(-i\lambda \hat{D})
\ \ ,
\label{eq:Iwasawa2}
\\
U\bigl(\exp[\theta(e^+-e^-)]\bigr)
& = &
\exp\bigl(-i\theta(\hat{H}_1-\hat{H}_2)\bigr)
\ \ .
\label{eq:Iwasawa3}
\ee
\end{subequations}
The operators in (\ref{eq:Iwasawa1}) 
and (\ref{eq:Iwasawa2}) 
act as
\begin{subequations}
\label{eq:iwaop-two-action}
\be
[ \exp (-i\mu\hat{H}_2)  \Psi ] (\boldsymbol{u},\boldsymbol{v})
& = &
\int \frac{d^q\boldsymbol{v}'}{(2\pi i\mu)^{q/2}}
\exp\left[\frac{i}{2}
\left(\frac{(\boldsymbol{v}-\boldsymbol{v}')^2}
{\mu}+\mu\boldsymbol{u}^2
\right) \right] 
\Psi(\boldsymbol{u},\boldsymbol{v}')
\ \ ,
\phantom{aaaa}
\label{eq:iwaop-two-action1}
\\
{[\exp (-i\lambda\hat{D}) \Psi]} (\boldsymbol{u},\boldsymbol{v})
& = &
\exp \! \left[\frac{\lambda}{2}(q-p) \right]
\Psi(e^{-\lambda}\boldsymbol{u},e^\lambda \boldsymbol{v})
\ \ .
\label{eq:iwaop-two-action2}
\ee
\end{subequations}
In (\ref{eq:Iwasawa3}) we have
$\hat{H}_1-\hat{H}_2 =
\hat{H}_{\boldsymbol{u}}^\mathrm{sho}-
\hat{H}_{\boldsymbol{v}}^\mathrm{sho}$,
where $\hat{H}_{\boldsymbol{u}}^\mathrm{sho}$ and
$\hat{H}_{\boldsymbol{v}}^\mathrm{sho}$ are the harmonic oscillator
Hamiltonians of unit mass and angular frequency in respectively
$\boldsymbol{u}$ and~$\boldsymbol{v}$. It follows that
$U(\exp[\theta(e^+-e^-)])$ is periodic in $\theta$ with period $2\pi$
when $p + q \equiv 0 \pmod2$ and with period $4\pi$ when $p + q \equiv
1 \pmod2$.  This means that the gauge group 
is $\SLtwor$ when $p + q \equiv 0 \pmod2$ and the double cover
of $\SLtwor$ when $p + q \equiv 1 \pmod2$. $U$~is isomorphic to the
$(p,q)$ oscillator representation of the double cover of $\SLtwor$
\cite{Howe} via the Fourier transform in~$\boldsymbol{v}$.

\subsection{Test space}
\label{subsec:test-space}

The next step is to introduce a linear space of test states
in~$\Haux$. The harmonic oscillator Hamiltonians in
$U\bigl(\exp[\theta(e^+-e^-)]\bigr)$
suggest that we make use of the harmonic
oscillator eigenstates in $\boldsymbol{u}$ and~$\boldsymbol{v}$,
\begin{equation}
\Psi_{ljmnk_uk_v} (\boldsymbol{u},\boldsymbol{v})
:=
u^l v^j e^{-\frac{1}{2}(u^2+v^2)}L_m^{\tilde{l}}(u^2)L_n^{\tilde{j}}(v^2)
Y_{lk_u}
\bigl({\theta}^{(u)}\bigr)
Y_{jk_v}
\bigl({\theta}^{(v)}\bigr)
\ \ ,
\label{eq:testfun}
\end{equation}
where $l$, $j$, $m$ and $n$ are non-negative integers,
$\tilde{l}$ and $\tilde{j}$ are defined by
\be
\tilde{l} := l+ (p-2)/2
\ \ ,
\hspace{7.5ex}
\tilde{j}:= j+ (q-2)/2 
\ \ ,
\label{eq:ltilde-jtilde}
\ee
$u := |\boldsymbol{u}|$, $v := |\boldsymbol{v}|$,
the $L$'s are the generalised Laguerre polynomials
\cite{arfken,magnusetal} 
and the $Y$'s are the spherical harmonics in the notation of 
section~\ref{sec:AQ}. These states satisfy
\be
&&
\hat{H}_{\boldsymbol{u}}^\mathrm{sho}\Psi_{ljmnk_uk_v}
=
E_u\Psi_{ljmnk_uk_v}
\ \ , \ \ \
E_u := 2m+l + (p/2)  = 2m+\tilde{l}+1
\ \ ,
\nn
\\
&&
\hat{H}_{\boldsymbol{v}}^\mathrm{sho}\Psi_{ljmnk_uk_v}
=
E_v\Psi_{ljmnk_uk_v}
\ \ , \ \ \
E_v := 2n+j + (q/2) = 2n+\tilde{j}+1
\ \ ,
\ee
and they are orthogonal in~$\Haux$,
\be
&&
\left(
\Psi_{ljmnk_uk_v}
\ ,
\Psi_{l' j' m' n'
k'_u k'_v}
\right)_\mathrm{aux}
\nn
\\
\noalign{\medskip}
&&
\qquad
=
\frac{
\Gamma\bigl(l + m + (p/2)\bigr)
\Gamma\bigl(j + n + (q/2)\bigr)
}
{4 \Gamma(m+1) \Gamma(n+1)}
\, 
\delta_{l l' }
\delta_{j j' }
\delta_{m m' }
\delta_{n n' }
\delta_{k_u k'_u}
\delta_{k_v k'_v}
\ \ .
\label{psis-ip-aux}
\ee
We set $\Phi_0 := \mathrm{span} \{ \Psi_{ljmnk_uk_v} \} =
\left\{ P(\boldsymbol u, \boldsymbol v)
\exp\left[-\tfrac12 (u^2 + v^2) \right]
\right\}$,
where
$P(\boldsymbol u, \boldsymbol v)$ is
an arbitrary polynomial in $\{ u_i \}$ and $\{ v_i \}$.
$\Phi_0$ is clearly dense in $\Haux$ and
mapped to itself by the quantum
constraints~(\ref{eq:con-op}).

Let $G$ denote the gauge group, and let
$dg$ be the (left and right) invariant Haar measure on~$G$. 
An $L^1$ function $h$ on $G$ defines on $\Haux$ the bounded
operator $\hat{h} := \int_G dg \, h(g) U(g)$, and the set of all such
operators generates an algebra~$\Agrouphat$. Starting with~$\Phi_0$,
we first take the closure under the algebra generated by $\{ U(g) \mid
g\in G \}$, then take the closure under~$\Agrouphat$, and adopt
the resulting space $\Phi$ as our test space. $\Phi$~is a dense linear
subspace of~$\Haux$, invariant under both $\Agrouphat$ and the algebra
generated by $\{ U(g) \mid g\in G \}$, and it hence satisfies the test
space postulates of~\cite{GM2}.

\subsection{Physical Hilbert space}
\label{subsec:group-aver}

We now construct a rigging map by averaging states in $\Phi$
over~$G$.

We define on $\Phi$ the sesquilinear form
\begin{equation}
(\phi_2,\phi_1)_\mathrm{ga}
:=
\int_G
dg \,
(\phi_2,U(g)\phi_1)_\mathrm{aux}
\ \ .
\label{eq:ga}
\end{equation}
We show in appendix~\ref{app:convergence}, 
Theorem~\ref{theorem:ga-convergence}, 
that the integral in
(\ref{eq:ga}) is absolutely convergent for all 
$\phi_1, \phi_2 \in \Phi$,
and
$(\cdot\,,\cdot )_\mathrm{ga}$ is hence well defined. 
We also show that
$(\cdot\,,\cdot )_\mathrm{ga}$ vanishes for $p+q \equiv 1
\pmod2$. For the rest of this subsection we take 
$p+q \equiv 0 \pmod2$. 

Let $\Phi^*$ be the algebraic dual of $\Phi$
and let $f[\phi]$ denote the
dual action of $f\in \Phi^*$ on $\phi \in \Phi$.
We define the
antilinear
map $\eta: \Phi \to \Phi^*$ by
\begin{equation}
\eta(\phi_1)[\phi_2]
:=
{(\phi_1 , \phi_2 )}_\mathrm{ga}
\ \ ,
\label{eq:GAM-eta}
\end{equation}
and we define on the image of $\eta$ the sesquilinear form
$(\cdot\,,\cdot )_\mathrm{RAQ}$ by
\begin{equation}
\bigl(\eta(\phi_1), \eta(\phi_2) \bigr)_\mathrm{RAQ}
:= \eta(\phi_2)[\phi_1]
\ \ .
\label{eq:raq-ip}
\end{equation}
We need to investigate whether 
the image of $\eta$ is nontrivial and whether
$(\cdot\,,\cdot )_\mathrm{RAQ}$ is positive definite.
If yes, 
$\eta$ is a rigging map and 
the physical Hilbert space is the Cauchy completion of the image of
$\eta$ in $(\cdot\,,\cdot)_\mathrm{RAQ}$. 

Note first that if
$\phi_i\in\Phi$ and $h_i \in L^1(G)$,
(\ref{eq:ga}) and (\ref{eq:GAM-eta})
imply \cite{GM2}
\begin{equation}
\eta( {\hat{h}}_1 \phi_1) [{\hat{h}}_2 \phi_2]
=
\overline{
\left(\int_G d g \, h_1(g) \right)
}
\left(\int_G d g \, h_2(g) \right)
\eta(\phi_1)[\phi_2]
\label{eq:eta-groupalg}
\end{equation}
and
$\eta(\phi_1) [ U(g) \phi_2]
=
\eta\bigl(U(g)\phi_1\bigr) [\phi_2]
= \eta(\phi_1) [\phi_2]$ for all~$g$. 
Hence it suffices to evaluate
$\eta(\phi_1) [\phi_2]$ for $\phi_1,\phi_2 \in \Phi_0$.

By Proposition \ref{proposition:fubini} 
in appendix~\ref{app:convergence}, 
Fubini's theorem implies that we can represent
the image of
$\eta$ as functions on 
$\BbbR^{p+q} = \{(\boldsymbol{u}, \boldsymbol{v})\}$,
acting on $\phi\in\Phi$ by
\begin{equation}
f[\phi] = \int 
d^p \boldsymbol{u} 
\, d^q \boldsymbol{v}
\, f (\boldsymbol u, \boldsymbol v) 
\phi(\boldsymbol{u}, \boldsymbol{v})
\ \ ,
\label{eq:dual-action}
\end{equation}
and evaluate $\eta$ by
\begin{equation}
\eta (\phi)
=
\overline{\int_G dg \, U(g) \phi}
\ \ ,
\label{ga-formal}
\end{equation}
where the integral is taken in the sense of pointwise convergence on
$\BbbR^{p+q}$. The value of (\ref{ga-formal}) can be read off from the
results in appendix D.1 of~\cite{LouRov}, by matching our 
(\ref{eq:UPsi-raw2}) to equation (D3) in~\cite{LouRov}. 
The result is 
\begin{equation}
\eta\bigl(\Psi_{ljmnk_uk_v}\bigr)
= 4 \pi^2 {(-1)}^{m}
\,
\delta_{mn}
\frac{\Gamma\bigl(l + m + (p/2)\bigr)}
{(2l + p - 2) \Gamma(m+1)}
\overline{ \Psi_{ljk_uk_v} }
\ \ ,
\label{eq:eta-Psi-explicit}
\end{equation}
where $\Psi_{ljk_uk_v}$ is as
in~(\ref{eq:physstates}). 
The action of $\overline{\Psi_{ljk_uk_v}}$
on $\Phi_0$ reads 
(\cite{magnusetal}, p.~244) 
\begin{equation}
\overline{ \Psi_{l'j' k'_u k'_v} }
\,
\bigl[ \Psi_{ljmnk_uk_v} \bigr]
=
(-1)^m
\delta_{2l+p,2j+q}
\delta_{l l' }
\delta_{j j' }
\delta_{m n}
\delta_{k_u k'_u}
\delta_{k_v k'_v}
\frac{\Gamma\bigl(l + m + (p/2)\bigr)}
{2\Gamma(m+1)}
\ \ .
\label{eq:rigging-explicit}
\end{equation}
Hence the image of $\eta$ is nontrivial and spanned by
$\bigl\{ \,
\overline{\Psi_{ljk_uk_v}}
\, \bigr\}$.
{}From~(\ref{eq:raq-ip}),
(\ref{eq:eta-Psi-explicit}) and (\ref{eq:rigging-explicit}) we find
\begin{equation}
\bigl(
\,
\overline{\Psi_{l'j'k'_uk'_v}} ,
\overline{\Psi_{ljk_uk_v}}
\,
\bigr)_{\mathrm{RAQ}} =
\frac{2l + p - 2}{8 \pi^2} 
\, 
\delta_{2l+p,2j+q}
\delta_{l l' }
\delta_{j j' }
\delta_{k_u k'_u}
\delta_{k_v k'_v}
\ \ .
\label{eq:eta-explicit}
\end{equation}
Hence $(\cdot\,,\cdot )_\mathrm{RAQ}$ is positive definite, $\eta$ is
a rigging map, and we have a physical Hilbert
space~$\Hraq$. 
The group averaging sesquilinear form on 
$\Phi_0$ reads
\be
(\Psi_{l'j'm'n'k_u'k_v'} , \Psi_{ljmnk_uk_v})_{\mathrm{ga}} 
&=& 
2 \pi^2 {(-1)}^{m + m'}
\,
\delta_{2l+p,2j+q}
\delta_{mn} \delta_{m'n'}
\delta_{l l' }
\delta_{j j' }
\delta_{k_u k'_u}
\delta_{k_v k'_v}
\nn
\\
&&
\times 
\frac{\Gamma\bigl(l + m' + (p/2)\bigr) \Gamma\bigl(l + m + (p/2)\bigr)}
{(2l + p - 2) \Gamma(m'+1) \Gamma(m+1)}
\ \ . 
\label{eq:ga-ip-explicit}
\ee
The uniqueness theorem of \cite{GM2} shows that 
every rigging map for our triple
$(\Haux,U,\Phi)$ is a multiple of the group averaging rigging
map~$\eta$.

The algebra $\Aphystar$ is represented on $\Haux$
by~(\ref{eq:q-obser}).
This representation leaves $\Phi$ invariant and commutes with~$U(g)$,
and the star-relation in this representation coincides with the
adjoint map on~$\Haux$. It follows that $\Hraq$ carries an antilinear
representation $\rho$
of~$\Aphystar$, such that the
star-relation coincides with the adjoint map on~$\Hraq$.
In the notation of~(\ref{eq:q-obser}),
\begin{equation}
\rho(\hat{\mathcal{O}}_{ij}) :
f \mapsto \overline{ \hat{\mathcal{O}}_{ij} \overline{f}}
\ \ .
\end{equation}
This shows that the algebraic quantisation set up in section
\ref{sec:AQ} yields a quantum theory anti-isomorphically embedded in
our group averaging quantum theory whenever $p\ge3$, $q\ge3$ and $p+q
\equiv 0 \pmod 2$, even though we were able to complete the algebraic
quantisation explicitly only for $p=q=3$. Apart from $p=q=3$, we do
however not know whether this quantum theory is the \emph{only\/} one
arising from the algebraic quantisation for 
$p\ge3$, $q\ge3$ and $p+q
\equiv 0 \pmod 2$.

\section{Refined algebraic quantisation for $p < 3$ or $q < 3$}
\label{sec:RAQ-low}

In section \ref{sec:RAQus} we assumed $p\ge3$ and $q\ge3$. We now
discuss refined algebraic quantisation for lower $p$ or~$q$. 
By interchange of
$\boldsymbol{u}$ and~$\boldsymbol{v}$, it suffices
to consider $p\le q$.

\subsection{$p =1$, $q > 3$}
\label{subsec:p=1,q>3}

When $p =1$ and $q > 3$, 
we define $\Haux$ and $\Phi$ as in 
section~\ref{sec:RAQus}. The 
$u_1$-dependence of the test states (\ref{eq:testfun}) can be
written in terms of Hermite polynomials as $H_{l + 2m}(u_1)
\exp(-\tfrac12 u_1^2)$ (\cite{magnusetal}, p.~240), but the notation in
(\ref{eq:testfun}) covers also $p=1$, 
the spherical harmonics on $S^0$ being 
as described in subsection~\ref{subsec:alg-gen-pq}. 
We drop the redundant index 
$k_u$ and write 
\begin{equation}
\phi_{ljmnk} (u_1,\boldsymbol{v})
:= 
\Psi_{ljmn0k}
= 
u^l v^j e^{-\frac{1}{2}(u^2+v^2)} 
L_m^{\tilde{l}}(u^2) 
L_n^{\tilde{j}}(v^2)
Y_{l}
\bigl({\theta}^{(u)}\bigr)
Y_{jk}
\bigl({\theta}^{(v)}\bigr)
\ \ ,
\label{eq:testfun-1q}
\end{equation}
where $l \in \{0,1\}$ and $\tilde{l} = l - \frac12$. 

As a preliminary, let $Y_{j0} \bigl({\theta}^{(v)}\bigr)$ denote the
zonal spherical harmonics, which depend only on $v_q/v$ and are given
by Gegenbauer polynomials~\cite{Bate}. The recursion relations for 
the 
Gegenbauer polynomials and the generalised Laguerre
polynomials \cite{magnusetal} 
allow an explicit computation of the action of 
$\hat{C}_{1q}$ on~$\phi_{ljmn0}$. We find 
\begin{subequations}
\label{eq:1q-C}
\be
\hat{C}_{1q} 
\phi_{0jmn0}
& = & 
-W_{qj}
\bigl[
(n+\tilde{j})\phi_{1,j-1,m-1,n,0}+(n+1)\phi_{1,j-1,m,n+1,0}
\bigr]
\nn
\\
& & 
+ W_{q, j+1}
\bigl(
\phi_{1,j+1,m-1,n-1,0}
+ 
\phi_{1,j+1,mn0}
\bigr)
\ \ , 
\label{eq:1q-C-raise}
\\
\hat{C}_{1q} 
\phi_{1jmn0}
& = & 
W_{qj}
\bigl[
(m+\textstyle{\frac12})
(n+\tilde{j})
\phi_{0,j-1,mn0}
+
(m+1)(n+1)\phi_{0,j-1,m+1,n+1,0}
\bigr]
\nn
\\
& & 
-W_{q,j+1}
\bigl[
(m+\textstyle{\frac12})\phi_{0,j+1,m,n-1,0}
+ 
(m+1)\phi_{0,j+1,m+1,n0}
\bigr]
\ \ , 
\label{eq:1q-C-lower}
\ee
\end{subequations}
where 
\be
W_{qj}
&:=& 
2\left[\frac{j(j+q-3)}{(2j + q -2)(2j + q - 4)}\right]^{1/2} 
 \ \ \ \text{for  $j>0$}
\ \ , 
\nn
\\
W_{q0}
&:=& 
0 
\ \ , 
\label{eq:1q-CW}
\ee
and any $\phi_{ljmn0}$ on the right-hand side 
with $m<0$ or $n<0$ is understood as zero. 

Now, by Theorem~\ref{theorem:ga-convergence}, the group averaging
converges in absolute value.  When $q$ is even, the
$\theta$-dependence in (\ref{eq:UPsi-explicit}) shows that the image
of $\eta$ is trivial. In the rest of this subsection we take $q$ odd
and show that the
image of $\eta$ is trivial also in this case. 

It suffices to show that
$(\phi_{l'j'm'n'k'},\phi_{ljmnk})_\mathrm{ga}$ vanishes. When
$l=l'=1$, we can proceed as in subsection
\ref{subsec:group-aver} and the result follows
from~(\ref{eq:ga-ip-explicit}). When $l=0$ or $l'=0$,
(\ref{eq:UPsi-explicit}) shows that it suffices to consider $(\phi_{0
j m'n'0},\phi_{0 j mn 0})_\mathrm{ga}$. The $\theta$-dependence in
(\ref{eq:UPsi-explicit}) shows that the integral over $\theta$ gives
zero unless $2m = 2n + j + (q - 1)/2$, and a similar observation with
$U(g)$ conjugated to act on the first argument shows that the integral
over $\theta$ gives zero unless $2m' = 2n' + j + (q - 1)/2$. When $q =
5 + 4a$, $a = 0,1,\ldots$, it therefore suffices to consider
$(\phi_{0, 2s,n'+s+a+1,n'}, \phi_{0, 2s,n+s+a+1,n})_\mathrm{ga}$,
where $s$, $n$ and $n'$ are non-negative integers and we have
suppressed the last index of the $\phi$'s, understood to take the
value zero. When $q = 3 + 4b$, $b = 1,2,\ldots$, it similarly suffices
to consider $(\phi_{0, 2s+1,n'+s+b+1,n'}, \phi_{0,
2s+1,n+s+b+1,n})_\mathrm{ga}$, where $s$, $n$ and $n'$ are
non-negative integers.

Let $q = 3 + 4b$, $b = 1,2,\ldots$. Recall that $\hat{C}_{1q}$ is
self-adjoint in $\Haux$ and commutes
with~$U(g)$. We compute
\be 
&&
W_{q,2s+1}
\bigl[
(n+s+b+{\textstyle\frac12}) 
(\phi_{0, 2s+1,n'+s+b+1,n'} , \phi_{0, 2s+1,n+s+b,n-1} )_\mathrm{ga}
\nn\\ 
&& 
\hspace{8ex}
+
(n+s+b+1) 
(\phi_{0, 2s+1,n'+s+b+1,n'} , \phi_{0, 2s+1,n+s+b+1,n} )_\mathrm{ga}
\bigr]
\nn\\ 
&& 
\hspace{3ex}
= \; 
- (\phi_{0, 2s+1,n'+s+b+1,n'} , 
\hat{C}_{1q}\phi_{1, 2s,n+s+b,n})_\mathrm{ga} 
\nn\\
&& 
\hspace{3ex}
= \; 
- (\hat{C}_{1q} \phi_{0, 2s+1,n'+s+b+1,n'} , 
\phi_{1, 2s,n+s+b,n})_\mathrm{ga} 
\nn\\
&& 
\hspace{3ex}
= \; 
0
\ \ , 
\label{eq:gaphiprime} 
\ee 
where the first equality follows from (\ref{eq:1q-C-lower})
and the last from 
(\ref{eq:1q-C-raise}) and~(\ref{eq:ga-ip-explicit}). 
By induction in~$n$, 
(\ref{eq:gaphiprime}) implies
$(\phi_{0, 2s+1,n'+s+b+1,n'} , \phi_{0, 2s+1,n+s+b+1,n}
)_\mathrm{ga}=0$. 

Let then $q = 5 + 4a$, $a = 0,1,\ldots$. An argument 
similar to (\ref{eq:gaphiprime}) shows that 
$(\phi_{0, 2s,n'+s+a+1,n'}, \phi_{0, 2s,n+s+a+1,n})_\mathrm{ga}$
vanishes for $s>0$. When $s=0$, we compute 
\be 
&&
W_{q1}
\bigl[
(n+a+{\textstyle\frac32}) ( n + 2a + {\textstyle\frac52}) 
(\phi_{0, 0 , n'+a+1,n'} , \phi_{0, 0 ,n+a+1,n} )_\mathrm{ga}
\nn\\ 
&& 
\hspace{8ex}
+
(n+a+2) ( n + 1) 
(\phi_{0, 0 , n'+a+1,n'}  , \phi_{0, 0 ,n+a+2,n+1} )_\mathrm{ga}
\bigr]
\nn\\ 
&& 
\hspace{3ex}
= \; 
(\phi_{0, 0 , n'+a+1,n'} , 
\hat{C}_{1q}\phi_{1, 1 ,n+a+1,n})_\mathrm{ga} 
\nn\\
&& 
\hspace{3ex}
= \; 
(\hat{C}_{1q} \phi_{0, 0 , n'+a+1,n'} , 
\phi_{1, 1 ,n+a+1,n})_\mathrm{ga} 
\nn\\
&& 
\hspace{3ex}
= \; 
0
\ \ , 
\ee 
where the last
equality follows from (\ref{eq:1q-C-raise}) and~(\ref{eq:ga-ip-explicit}).
By induction in~$n$, it therefore suffices to consider 
$(\phi_{0, 0 , n'+a+1,n'} , \phi_{0, 0 ,a+1,0} )_\mathrm{ga}$. A
similar argument in $n'$ shows that it suffices to consider 
$(\phi_{0, 0 , a+1,0} , \phi_{0, 0 ,a+1,0} )_\mathrm{ga}$.

In $(\phi_{0, 0 , a+1,0} , U(g) \phi_{0, 0 ,a+1,0} )_\mathrm{aux}$, we
use (\ref{eq:UPsi-explicit}) and perform the elementary integration
over~$\boldsymbol{v}$. We then integrate over $G$ in the Haar
measure $dg = \tfrac12 dz\,d\mu\,d\theta$. The integration over
$\theta$ is elementary. Changing the variables in the inner integral
from $u$ to $y := u^2/z$ and in the outer integral from $\mu$ to 
$t := \mu z/(z+1)$, 
we find that 
$(\phi_{0, 0 , a+1,0} , \phi_{0, 0 ,a+1,0})_\mathrm{ga}$ 
equals a numerical constant times 
\be
&&
\int_0^\infty dz \, 
\frac{z^{a+\frac12}}{(z+1)^{2a + \frac32}}
\int_{-\infty}^\infty \frac{dt}{(1+it)^{2a + \frac52}}
\times 
\nn
\\
\noalign{\medskip}
&&
\ \ \times 
\int_0^\infty
dy \, y^{-\frac12} \, L^{-\frac12}_{a+1} (zy)  \, L^{-\frac12}_{a+1} (y) 
\exp \bigl[ - \tfrac12 (z+1) (1 - it) y \bigr] 
\ \ . 
\label{eq:1q-zero-1} 
\ee
We interchange the order of the $dt$ and $dy$ integrals
in~(\ref{eq:1q-zero-1}), 
justified by the absolute convergence of the double integral, and
perform the $dt$ integral as a contour integral, 
finding that (\ref{eq:1q-zero-1}) equals 
a numerical constant times 
\begin{equation}
\int_0^\infty dz \, z^{a+\frac12}
\int_0^\infty
dy \, y^{2a+1} 
L^{-\frac12}_{a+1} (zy)  \, L^{-\frac12}_{a+1} (y)
\exp [ - (z+1) y ]
\ \ . 
\label{eq:1q-zero-2} 
\end{equation}
In (\ref{eq:1q-zero-2}) we  
interchange the order of the $dz$ and $dy$ integrals, 
justified by the absolute convergence of the double integral. Changing
the variable in the new inner integral from $z$ to $x := zy$, we
obtain
\begin{equation}
\int_0^\infty
dy \, 
y^{a - \frac12}  \, L^{-\frac12}_{a+1} (y) \, e^{-y} 
\int_0^\infty dx \, 
x^{a+\frac12}  \, L^{-\frac12}_{a+1} (x) \, e^{-x} 
\ \ . 
\label{eq:1q-zero-final} 
\end{equation}
The integrals in (\ref{eq:1q-zero-final}) have factorised, and the
integral over $y$ vanishes by the orthogonality of the generalised
Laguerre polynomials~\cite{magnusetal}.

\subsection{$p =1$, $q = 3$}
\label{subsec:p=1,q=3}

When $p=1$ and $q=3$, we define $\Haux$ as in
section~\ref{sec:RAQus}. With $\Phi$ defined as in
section~\ref{sec:RAQus}, the integral in (\ref{eq:ga}) is not
absolutely convergent for $l=j=0$, and we have not found a 
weaker unambiguous sense of convergence. The $\theta$-dependence in
(\ref{eq:UPsi-explicit}) however suggests that if group averaging can
be made well-defined, it should annihilate states with $l=j=0$. We
shall achieve this by suitably modifying the 
test space.

Dropping the redundant index $k_u$, we introduce the states
$\phi_{ljmnk}$ by (\ref{eq:testfun-1q}) with $q=3$. We define 
$\Phi_0^\mathrm{mod} 
:= 
\mathrm{span}
\bigl(
\{ \phi_{ljmnk} \mid l+j>0 \} \cup \{ \psi_{mn} \}
\bigr)$, where 
\begin{equation}
\psi_{mn}
:= 
\textstyle{\frac{2}{\sqrt{3}}}
\bigl[
(m+\textstyle{\frac12})(n+\textstyle{\frac32})\phi_{00mn0}
+
(m+1)(n+1)\phi_{00,m+1,n+1,0}
\bigr]
\ \ . 
\label{eq:13-psi-def}
\end{equation}
Using the basis (\ref{eq:q-obser}) of~$\Aphystar$, properties of
the spherical harmonics on $S^2$ 
\cite{Bate,arfken} and properties of the generalised 
Laguerre polynomials~\cite{magnusetal}, 
it can be verified  
that $\Phi_0^\mathrm{mod}$ is invariant
under~$\Aphystar$. 
In particular, formulas 
(\ref{eq:1q-C})
and 
(\ref{eq:1q-CW}) 
hold with $q=3$, 
implying 
\begin{subequations}
\be 
\hat{C}_{13}\phi_{11mn0}
&=& \psi_{mn}
-\textstyle{\frac{4}{\sqrt{15}}}
\bigl[
(m+\textstyle{\frac12})\phi_{0,2,m,n-1,0}
+
(m+1)\phi_{0,2,m+1,n,0}
\bigr]
\ \ , 
\label{eq:13-C-spe-low}
\\
\hat{C}_{13}\psi_{mn} 
& = & 
\textstyle{\frac43}
(m+\textstyle{\frac12})
(n+\textstyle{\frac32})
( \phi_{11,m-1,n-1,0} + \phi_{11mn0} )   
\nn
\\
& & + \textstyle{\frac43} 
(m+1)(n+1)( \phi_{11mn0} + \phi_{11,m+1,n+1,0} )
\ \ . 
\label{eq:13-C-spe-raise}
\ee
\end{subequations}

We claim that $\Phi_0^\mathrm{mod}$ is dense
in~$\Haux$. If this were not the case, there would exist a nonzero
vector $y = \sum_{ljmnk} a_{ljmnk} \phi_{ljmnk} \in \Haux$ that is
orthogonal to all vectors in $\Phi_0^\mathrm{mod}$. 
By (\ref{psis-ip-aux}), orthogonality with each
$\phi_{ljmnk}$ with $l+j>0$ implies $a_{ljmnk}=0$ for $l+j>0$.
By (\ref{psis-ip-aux}) and~(\ref{eq:13-psi-def}), 
orthogonality with each 
$\psi_{mn}$ implies 
$a_{00mn0} + a_{00,m+1,n+1,0} =0$, from which 
(\ref{psis-ip-aux}) further shows that 
$y$ has
finite norm only if $y$ is the zero vector. 
Hence $\Phi_0^\mathrm{mod}$ is dense
in~$\Haux$. 

Following section \ref{sec:RAQus} with $\Phi_0$ replaced
by~$\Phi_0^\mathrm{mod}$, we first take the closure of 
$\Phi_0^\mathrm{mod}$  
under the algebra generated by $\{ U(g) \mid g\in G \}$, 
then take the
closure under~$\Agrouphat$, and adopt the resulting space 
$\Phi^\mathrm{mod}$ as our test space. 
$\Phi^\mathrm{mod}$~is a dense linear subspace of~$\Haux$,
invariant under $\Aphystar$, 
$\Agrouphat$ and the algebra generated by $\{
U(g) \mid g\in G \}$, 
and satisfies hence the test space postulates of~\cite{GM2}.
We show in appendix~\ref{app:convergence}, 
Theorem~\ref{theorem:ga-13-convergence}, that 
the integral in (\ref{eq:ga})
converges in absolute value for all
$\phi_1,\phi_2 \in \Phi^\mathrm{mod}$. 

To evaluate $(\phi_2,\phi_1)_\mathrm{ga}$
on~$\Phi^\mathrm{mod}$, 
it suffices to consider 
$\phi_1,\phi_2 \in \Phi_0^\mathrm{mod}$. 
When both $\phi_1$ and $\phi_2$ have $l=1$, we can proceed as in
subsection~\ref{subsec:group-aver}, arriving at
(\ref{eq:dual-action})--(\ref{eq:ga-ip-explicit}). When $\phi_1$ and
$\phi_2$ have differing values of $l$, $j$ or~$k$, (\ref{eq:UPsi-explicit})
shows that $(\phi_2,\phi_1)_\mathrm{ga}$ vanishes. What remains is 
$(\phi_{0jm'n'k},\phi_{0jmnk})_\mathrm{ga}$ with $j>0$ and 
$(\psi_{m'n'},\psi_{mn})_\mathrm{ga}$. The vanishing of the former
follows as in subsection~\ref{subsec:p=1,q>3}, noting that 
(\ref{eq:gaphiprime}) holds also for $b=0$. For the latter, 
we use~(\ref{eq:13-C-spe-low}), 
the self-adjointness of $\hat{C}_{13}$ on $\Haux$ 
and the vanishing of 
$(\psi_{m'n'},\phi_{0jmn0})_\mathrm{ga}$ for $j>0$ and compute
\be 
(\psi_{m'n'},\psi_{mn})_\mathrm{ga} 
& = &
(\psi_{m'n'},\hat{C}_{13} \phi_{11mn0})_\mathrm{ga} 
\nn
\\ 
& = &
(\hat{C}_{13} \psi_{m'n'}, \phi_{11mn0})_\mathrm{ga} 
\nn
\\ 
& = & 0 
\ \ , 
\ee 
where the last equality follows from 
(\ref{eq:13-C-spe-raise}) and~(\ref{eq:ga-ip-explicit}). 

The evaluation of $(\phi_2,\phi_1)_\mathrm{ga}$ is now complete. The only
nonzero contribution comes from states with $l=1$, in which case
formulas (\ref{eq:dual-action})--(\ref{eq:ga-ip-explicit}) hold. The image
of $\eta$ is one-dimensional, spanned by 
$\bigl\{ \,
\overline{\Psi_0} 
\, 
\bigr\}$, 
where $\Psi_0$ is the state (\ref{eq:physstates}) with $l=1$ and $j=0$ 
and reads explicitly (\cite{Bate}, Section~7.11)
$\Psi_{0} = v^{-1}\sin(u_1v)$. The inner product (\ref{eq:eta-explicit}) is
positive definite, and we obtain a one-dimensional physical Hilbert
space~$\Hraq$. 

As $\Phi_0^\mathrm{mod}$ is invariant under~$\Aphystar$, $\Hraq$
carries an antilinear representation of~$\Aphystar$. A~direct
calculation shows that all operators in this representation
annihilate~$\overline{\Psi_0}$, and the representation is trivial. The
quantum theory found in algebraic quantisation in subsection
(\ref{subsec:alg-gen-pq}) is thus anti-isomorphically embedded in the
group averaging quantum theory.

\subsection{$p =1$, $q = 2$}

When $p=1$ and $q=2$, and $\Haux$ and $\Phi$ are as in
section~\ref{sec:RAQus}, the integral in (\ref{eq:ga}) is not
absolutely convergent for $l=j=0$. It may be possible to modify the
$l=j=0$ sector of $\Phi$ as in subsection \ref{subsec:p=1,q=3}
above, but as now $p+q \equiv 1 \pmod2$, any test space built
from linear combinations of the harmonic oscillator eigenfunctions
will give an $\eta$ with trivial image.

\subsection{$p=q=1$}
\label{subsec:p=q=1}

When $p=q=1$, and $\Haux$ and $\Phi$ are as in 
section~\ref{sec:RAQus}, the integral 
in (\ref{eq:ga}) is not absolutely convergent for $l=0$ or $j=0$ and is
unambiguously divergent for example for $\phi_1 = \phi_2 =
\Psi_{0000}$. 

We attempt to cure the divergence by modifying the zero angular
momentum sector. For technical simplicity, we choose at the outset
to work with 
states that are symmetric under $(u_1,v_1) \mapsto (-u_1,-v_1)$.

Let 
$\Hauxsym \subset \Haux$
be the Hilbert subspace of vectors symmetric under
$(u_1,v_1) \mapsto (-u_1,-v_1)$. Dropping the 
redundant indices 
$k_u$ and~$k_v$, we write 
\begin{equation}
\phi_{lmn} (u_1,v_1)
:= 
\Psi_{llmn00}
= 
u^l v^l e^{-\frac{1}{2}(u^2+v^2)} 
L_m^{\tilde{l}}(u^2) 
L_n^{\tilde{l}}(v^2)
Y_{l}
\bigl({\theta}^{(u)}\bigr)
Y_{l}
\bigl({\theta}^{(v)}\bigr)
\ \ ,
\label{eq:testfun-11}
\end{equation}
where $l \in \{0,1\}$ and $\tilde{l} = l - \frac12$. 
$\{ \phi_{lmn} \}$~is clearly an orthogonal basis for~$\Hauxsym$. 

Let $\Phi_0^\mathrm{s} := \mathrm{span} \{ \psi_{mn}, \phi_{1mn} \}$, 
where 
\begin{equation}
\psi_{mn}
:= 
2
\bigl[
(m + \textstyle{\frac12}) 
(n + \textstyle{\frac12}) 
\phi_{0mn}
+ 
(m + 1) 
(n + 1) 
\phi_{0,m+1,n+1}
\bigr]
\ \ . 
\label{eq:psimn-def}
\end{equation}
We then find (\cite{magnusetal}, p.~241) 
\begin{subequations}
\label{eq:Phinoughtbasis}
\be
\hat{C}_{11} \psi_{mn}
&=&
4
\bigl[
(m + \textstyle{\frac12}) 
(n + \textstyle{\frac12}) 
(\phi_{1mn} + \phi_{1,m-1,n-1} )
\nn
\\
&&
\quad
+ 
(m + 1) 
(n + 1) 
(\phi_{1mn} + \phi_{1,m+1,n+1})
\bigr]
\ \ , 
\label{eq:Phinoughtbasis1}
\\
\hat{C}_{11} \phi_{1mn}
&=& 
\psi_{mn}
\ \ , 
\label{eq:Phinoughtbasis2}
\ee
\end{subequations}
where $\hat{C}_{11}$ (\ref{eq:q-obser}) is the single
generator of~$\Aphystar$. Hence $\Phi_0^\mathrm{s}$ is invariant
under~$\Aphystar$, and it can be shown as in subsection
\ref{subsec:p=1,q=3} that $\Phi_0^\mathrm{s}$ is dense
in~$\Hauxsym$. We build from $\Phi_0^\mathrm{s}$ a test
space $\Phi^\mathrm{s}$ satisfying the postulates of \cite{GM2} as in
subsection~\ref{subsec:p=1,q=3}. 
The integral in (\ref{eq:ga}) then 
converges in absolute value for all
$\phi_1,\phi_2 \in \Phi^\mathrm{s}$: 
The proof is a verbatim adaptation 
of that of Theorem~\ref{theorem:ga-13-convergence}. 

We need to evaluate 
$(\phi_2,\phi_1)_\mathrm{ga}$ on~$\Phi^\mathrm{s}$. It suffices 
to consider 
$\phi_1,\phi_2 \in \Phi_0^\mathrm{s}$. 
Clearly $(\psi_{m'n'},\phi_{1mn})_\mathrm{ga} =0$. For 
$(\phi_{1m'n'},\phi_{1mn})_\mathrm{ga}$ we proceed as in
subsection \ref{subsec:group-aver} and arrive at 
(\ref{eq:dual-action})--(\ref{eq:ga-ip-explicit}), 
the last of which
reads
\begin{equation}
(\phi_{1m'n'},\phi_{1mn})_\mathrm{ga}
= 
2\pi^2 (-1)^{m+m'} \delta_{mn} \delta_{m'n'} 
\frac{
\Gamma\bigl(m + \frac32 \bigr) 
\Gamma\bigl(m' + \frac32 \bigr) 
}
{\Gamma(m+1) \Gamma(m'+1)}
\ \ . 
\label{eq:11-phi-phi}
\end{equation}
To find
$(\psi_{m'n'},\psi_{mn})_\mathrm{ga}$, 
we use the self-adjointness of
$\hat{C}_{11}$ on $\Hauxsym$ and compute 
\be
(\psi_{m'n'},\psi_{mn})_\mathrm{ga}
&=&
(\psi_{m'n'}, \hat{C}_{11}\phi_{1 mn})_\mathrm{ga}
\nn
\\
&=&
(\hat{C}_{11} \psi_{m'n'}, \phi_{1 mn})_\mathrm{ga}
\nn
\\
&=&
- 2\pi^2 (-1)^{m+m'} \delta_{mn} \delta_{m'n'} 
\frac{
\Gamma\bigl(m + \frac32 \bigr) 
\Gamma\bigl(m' + \frac32 \bigr) 
}
{\Gamma(m+1) \Gamma(m'+1)}
\ \ , 
\label{eq:11-psi-psi}
\ee
where the first equality follows from (\ref{eq:Phinoughtbasis2}) and the
last one from (\ref{eq:Phinoughtbasis1}) and~(\ref{eq:11-phi-phi}). 

We see that $(\cdot\,,\cdot )_\mathrm{ga}$ is an \emph{indefinite\/}
sesquilinear form. Hence the map $\eta$ defined by 
(\ref{eq:GAM-eta}) is not a rigging map and we do not recover a
Hilbert space. The indefiniteness of $(\cdot\,,\cdot )_\mathrm{ga}$
further 
implies, by the uniqueness theorem of~\cite{GM2}, 
that the triple $(\Hauxsym,U,\Phi^\mathrm{s})$ admits no rigging maps. 

The image of $\eta$ 
is two-dimensional, 
spanned by 
$\bigl\{ \,
\overline{\Psi_{00}} 
\, , 
\overline{\Psi_{11}} 
\, 
\bigr\}$, 
where $\Psi_{00}$ and $\Psi_{11}$ are given by (\ref{eq:physstates}) 
and read explicitly (\cite{Bate}, Section~7.11) 
\begin{subequations}
\be
\Psi_{00} 
&=& 
\frac{1}{\sqrt{2\pi}}  \cos(u_1 v_1) 
\ \ ,
\\
\Psi_{11} 
&=& 
\frac{1}{\sqrt{2\pi}}  \sin(u_1 v_1) 
\ \ . 
\ee
\end{subequations}
The manifestly indefinite 
sesquilinear form (\ref{eq:raq-ip}) on the image of $\eta$
is given by~(\ref{eq:eta-explicit}). 
The representation of 
$\Aphystar$ induced on the image of 
$\eta$ by (\ref{eq:dual-action}) is
anti-isomorphic to the 
representation obtained in subsection \ref{subsec:alg-gen-pq} on the
solution space to the algebraic quantisation constraints.

\subsection{$p =2$, $q >2$}
\label{subsec:p=2,q>2}

When $p =2$ and $q > 2$, we define $\Haux$ and $\Phi$ as in 
section~\ref{sec:RAQus}. Theorem \ref{theorem:ga-convergence}
in appendix \ref{app:convergence} shows that the group
averaging converges in absolute value. 

When $q$ is odd, the $\theta$-dependence in
(\ref{eq:UPsi-explicit}) shows that the 
image of $\eta$ is trivial. 

Suppose then that $q$ is even. 
When $l>0$ and $l'>0$, we arrive at equations 
(\ref{eq:dual-action})--(\ref{eq:ga-ip-explicit})
as in subsection~\ref{subsec:group-aver}. When $l=0$ or $l'=0$, it can be
shown that $(\Psi_{l' j' m'n' k_u'
k_v'} ,
\Psi_{l j mn k_u k_v})_{\mathrm{ga}}$ vanishes: The arguments follow those
in subsection \ref{subsec:p=1,q>3} so closely that we will not spell them
out here. This means that equations 
(\ref{eq:dual-action})--(\ref{eq:ga-ip-explicit}) hold for all values of
the indices in the sense that terms involving
$\delta_{2l+1,2j+q}$ for
$l=0$ are understood to vanish. Hence the situation is similar to that for
$p\ge3$, $q\ge3$ and $p+q \equiv 0 \pmod2$ in section~\ref{sec:RAQus}. The
image of $\eta$ is nontrivial, $(\cdot\,,\cdot )_\mathrm{RAQ}$ is positive
definite, $\eta$ is a rigging map, and the representation of $\Aphystar$ on
the physical Hilbert space is as described at the end of
subsection~\ref{subsec:group-aver}.

\subsection{$p =q =2$}

The case $p =q =2$ was analysed in~\cite{LouRov}. Group averaging does
not converge on the test space of 
section~\ref{sec:RAQus}, but the $l=j=0$ sector of the test space 
can be modified so that group averaging converges and the physical 
observable algebra includes~$\Aphystar$. The physical Hilbert
space decomposes into a direct sum of four Hilbert subspaces, each of
them carrying a distinct representation of~$\Aphystar$.

\section{Discussion}
\label{sec:discussion}

We have discussed the quantisation of a constrained system
with unreduced phase space
$\BbbR^{2(p+q)}$, classical gauge group $\SLtwor$ and a distinguished
$\mathfrak{o}(p,q)$ algebra of classical observables. 
We employed refined algebraic 
quantisation, using group averaging on an auxiliary Hilbert
space to find the
inner product on the physical Hilbert space. We took care to select the
quantisation input so that when a quantum
theory is recovered, the classical 
$\mathfrak{o}(p,q)$ algebra gets promoted
into an operator algebra represented on the physical Hilbert space. 

When $p\ge2$, $q\ge2$, $p+q >4$ and $p+q \equiv 0 \pmod2$, we found a
quantum theory with a nontrivial representation of the $\mathfrak{o}(p,q)$
observables. For $p=q=2$, a similar result was obtained
in~\cite{LouRov}. For $(p,q) = (1,3)$ or 
$(3,1)$, we found a quantum theory with a one-dimensional Hilbert space and
a trivial representation of the $\mathfrak{o}(p,q)$ observables. For other
values of $p$ and $q$ we found no quantum theory. 

We also discussed Ashtekar's algebraic quantisation, solving first the
quantum constraints without an inner product and then promoting the
classical $\mathfrak{o}(p,q)$ algebra into operators whose
star-relations determine the physical inner product. For all values of
$p$ and $q$ for which group averaging gave a quantum theory, 
algebraic quantisation gave a quantum theory that is
(anti-)isomorphically embedded in the group averaging theory. For
$p=q=3$, we showed that 
this algebraic quantisation theory is unique. 

With both algebraic quantisation and group averaging, qualitative
changes emerged depending on whether $p$ and $q$ are less than, equal
to, or greater than~2. This could be expected from the properties of
the classical reduced phase space: The reduced phase space contains a
symplectic manifold when and only when $\min(p,q) \ge2$, and this
symplectic manifold is connected when and only when $\min(p,q)
\ge3$. However, a phenomenon not expected on classical grounds was
that neither algebraic quantisation nor group averaging gave a quantum
theory for $p+q \equiv 1 \pmod2$. The technical reason was that
both quantisation schemes represented the $\mathfrak{o}(p) \oplus
\mathfrak{o}(q)$ subalgebra of $\mathfrak{o}(p,q)$ by integer-valued
rather than half-integer-valued angular momenta. Obtaining quantum
theories for $p+q \equiv 1 \pmod2$ by some `fermionic' modification
might be an interesting challenge.

For $p=q=1$, both algebraic quantisation and group averaging failed to
give a quantum theory, for closely related reasons. Algebraic quantisation
led to a two-dimensional vector space of solutions to the constraints, but
requiring the $\mathfrak{o}(1,1)$ generator to be symmetric forced
the sesquilinear form on this vector space 
to be indefinite. In group averaging, a
judicious choice of the test space ensured convergence of the
averaging and the inclusion of the 
$\mathfrak{o}(1,1)$ generator in the would-be physical 
observable algebra, but the outcome was 
the same indefinite sesquilinear form on the same 
two-dimensional vector space as in 
algebraic quantisation. It is not clear whether the case $p=q=1$ has
physical interest, especially as the reduced phase space consists of just
three points, non-Hausdorff close to each other, but from the mathematical
point of view this provides the first example known to us where group
averaging fails to produce a Hilbert space owing to indefiniteness of
the would-be inner product. As the uniqueness theorem of
\cite{GM2} does not assume  positive definiteness, the theorem is
applicable here and  implies that our test space admits no rigging maps. 

We assumed throughout $p\ge1$ and $q\ge1$. If either $p$ or $q$
vanishes, the action (\ref{eq:act1}) still defines a classical theory,
but the reduced phase space then consists of a single point. Algebraic
quantisation in the representation of section \ref{sec:AQ} gives no
solutions to the constraints, and when group averaging based on the
harmonic oscillator eigenstates converges, it gives an
identically-vanishing sesquilinear form owing to the
$\theta$-dependence in $U(g)\Psi$~(\ref{eq:UPsi-explicit}).

Finally, one would like to characterise the
representations of $\mathfrak{o}(p,q)$ on our physical Hilbert
spaces in terms of invariants~\cite{AMMP}, 
as done in \cite{LouRov,MRT,TruSL} for
$p=q=2$. The value of the quadratic Casimir operator
can be read off from~(\ref{eq:pq33-quantum-casimir}). 
As our representation of the gauge group on the
auxiliary Hilbert space is isomorphic to the oscillator representation 
of $\SLtwor$~\cite{Howe}, the joint representation theory
of the dual pair $\bigl(\mathrm{O}(p,q),
\SLtwor\bigr)$ \cite{Howe-dual,PaulTan} may be useful with this
question.

\section*{Acknowledgements}

We thank 
John Barrett, 
Marco Mackaay, 
Don Marolf 
and 
Jacek Wi\'sniewski 
for discussions and correspondence, 
Bianca Dittrich for correspondence and bringing 
\cite{BarsKounnas,Bars98,Bars00,Bars01,Dittrich-pc} to our attention, 
and John Baez for 
bringing 
\cite{AMMP} to our attention. 
A.M. was
supported by a CONACYT (Mexico) Postgraduate Fellowship and an ORS
Award to the University of Nottingham.

\appendix

\section{Appendix: $\SLtwor$}
\label{app:SLtwor}

In this appendix we collect some relevant properties of $\SLtwor$. The
notation follows~\cite{Howe}.

$\SLtwor$ consists of real $2\times2$ matrices with unit
determinant. The Lie algebra $\sltwor$ is spanned by the matrices
\begin{equation}
h :=
\left(
\begin{array}{cc}
1 & 0
\\
0 & -1
\end{array}
\right)
\ \ , \ \
e^+ :=
\left(
\begin{array}{cc}
0 & 1
\\
0 & 0
\end{array}
\right)
\ \ , \ \
e^- :=
\left(
\begin{array}{cc}
0 & 0
\\
1 & 0
\end{array}
\right)
\ \ ,
\end{equation}
whose commutators are
\begin{eqnarray}
\left[ h \, , \,  e^+  \right]
& = &
2e^+
\ \ ,
\nonumber
\\
\left[ h \, , \,  e^-  \right]
& = &
-2e^-
\ \ ,
\nonumber
\\
\left[ e^+ \, , \, e^- \right]
& = &
h
\ \ .
\label{eq:sltwo-basiscomm}
\end{eqnarray}

Elements of $\SLtwor$ have the unique Iwasawa decomposition
\begin{equation} 
g
=
\exp(\mu e^-)
\exp(\lambda h)
\exp[\theta(e^+ - e^-)]
\ \ , 
\label{eq:iwasawa-decomp1}
\end{equation} 
or explictly 
\begin{equation} 
g
=
\left(\begin{array}{cc}
1   & 0 \\
\mu & 1
\end{array}\right)
\left(\begin{array}{cc}
e^\lambda & 0 \\
0         & e^{-\lambda}
\end{array}\right)
\left(\begin{array}{cc}
\cos\theta  & \sin\theta \\
-\sin\theta & \cos\theta
\end{array}\right)
\ \ ,
\label{eq:iwasawa-decomp2}
\end{equation}
where
$\mu\in\BbbR$, $\lambda\in\BbbR$ and
$0\le\theta<2\pi$.
The unique Iwasawa decomposition of the universal covering group of
$\SLtwor$ is given by (\ref{eq:iwasawa-decomp1}) with
$-\infty<\theta<\infty$, and that of the double cover by
$0\le\theta<4\pi$.
The
left and right invariant Haar measure reads
$dg = e^{2\lambda}\,d\lambda\,d\mu\,d\theta$.

\section{Appendix: Linear independence of the constraints}
\label{app:lin-ind}

In this appendix we show that 
the gradients of the constraints 
are all vanishing on~$\barGammanought$,
linearly dependent but not all vanishing on~$\barGammaex$, and linearly
independent on~$\barGammareg$.

{}From~(\ref{eq:const}), the gradients of the constraints read
\be
dH_1&=&\sum_i (p_idp_i-v_idv_i)
\ \ ,
\nn
\\
dH_2&=&\sum_i (\pi_id\pi_i-u_idu_i)
\ \ ,
\nn
\\
dD&=&\sum_i (u_idp_i+p_idu_i-\pi_idv_i-v_id\pi_i)
\ \ .
\label{eq:grad}
\ee
For $\alpha,\beta,\gamma \in \BbbR$, the equation
$\alpha dH_1+\beta dH_2+ \gamma dD=0$
is equivalent to 
\be
&&
\gamma \boldsymbol{u} + \alpha \boldsymbol{p}
= \boldsymbol{0}
=
- \beta \boldsymbol{u} + \gamma\boldsymbol{p}
\ \ ,
\nn
\\
&&
\gamma \boldsymbol{\pi} + \alpha\boldsymbol{v}
= \boldsymbol{0}
=
 - \beta\boldsymbol{\pi} + \gamma \boldsymbol{v}
\ \ .
\label{eq:contra}
\ee

$\barGammanought$ is clearly the 
set where the gradients of all the constraints vanish. 

On~$\barGammaex$, we saw in section \ref{sec:classical} that each
point can be brought to the form (\ref{eq:Mex-char}) by a gauge
transformation (\ref{eq:gauge-transf}) with some $g\in\SLtwor$. Given
such a~$g$, (\ref{eq:contra}) is satisfied by $\alpha = (g_{12})^2$,
$\beta = - (g_{11})^2$ and $\gamma = g_{11} g_{12}$, where at least
one of $\alpha$ and $\beta$ must be nonvanishing since $\det(g)\ne0$.
Hence the gradients of the constraints are
linearly dependent on~$\barGammaex$.

On~$\barGammareg$, the pair $(\boldsymbol{u},\boldsymbol{p})$ (as well
as the pair $(\boldsymbol{v},\boldsymbol{\pi})$)
is linearly independent, 
and (\ref{eq:contra}) implies
$\alpha=\beta=\gamma=0$. Hence the gradients of the constraints are
linearly independent on~$\barGammareg$.

\section{Appendix: Separation of $\Mreg$ by $\Aclass$}
\label{app:separation}

In this appendix we show that the classical observable algebra
$\Aclass$ separates~$\Mreg$. 
We assume $p\ge2$ and $q\ge2$, which is necessary
and sufficient for $\Mreg$ to be nonempty. The case $p=q=2$ was
treated in~\cite{LouRov,MRT,TruSL}.

Let $\mathcal{M}_i$, $1 \le i \le p$, be the subset of $\Mreg$ whose
points have a representative in $\barGammareg$ satisfying the gauge
conditions (\ref{eq:Mreg-char}) with $u_i^2 + p_i^2 >0$.  It follows
that $\mathcal{M}_i \cap
\mathcal{M}_j \ne \emptyset$ for all $i$ and $j$ and
\begin{equation}
\Mreg =
\bigcup_{i=1}^p \mathcal{M}_i
\ \ .
\end{equation}


\begin{lemma}
\label{lemma:Mi-distinct}
Let $\tilde{a}\in\Mreg$ and $\tilde{b}\in\Mreg$ such that there is no
$\mathcal{M}_i$ containing both $\tilde{a}$ and~$\tilde{b}$. Then
$\Aclass$ separates $\tilde{a}$ and~$\tilde{b}$.
\end{lemma}

\myproof
Let $a =
(\boldsymbol{u},\boldsymbol{p},
\boldsymbol{v},\boldsymbol{\pi}) 
\in \barGammareg$
and $b =
(\boldsymbol{u}',\boldsymbol{p}',
\boldsymbol{v}',\boldsymbol{\pi}') 
\in
\barGammareg$ be representatives of respectively
$\tilde{a}$ and~$\tilde{b}$, each satisfying~(\ref{eq:Mreg-char}).
As the pair $(\boldsymbol{u},\boldsymbol{p})$ is
linearly independent, there exist $i\ne j$ such that $u_i p_j - u_j
p_i \ne0$. It follows that
$\tilde{a} \in \mathcal{M}_i \cap \mathcal{M}_j$. By
assumption then $\tilde{b} \notin \mathcal{M}_i \cup \mathcal{M}_j$,
which implies $u'_i = p'_i = u'_j = p'_j =
0$. Hence 
$A_{ij}(\tilde{a}) =  u_i p_j - u_j p_i \ne0$
but
$A_{ij}(\tilde{b}) =  u'_i p'_j - u'_j p'_i =0$, 
which shows that 
the observable $A_{ij}$ (\ref{eq:obser})
separates $\tilde{a}$
and~$\tilde{b}$. 
$\blacksquare$

\myremark
Repeating the proof with $\tilde{a}$ and
$\tilde{b}$ interchanged shows that points satisfying the
conditions of Lemma \ref{lemma:Mi-distinct} exist
only for $p \ge 4$.

\begin{theorem}
\label{theorem:Mreg-separation}
$\Aclass$ separates~$\Mreg$.
\end{theorem}

\myproof
By Lemma~\ref{lemma:Mi-distinct}, it suffices to consider
individually each~$\mathcal{M}_k$.

{}From now on let $\mathcal{M}_k$ be fixed. We saw in subsubsection 
\ref{subsubsec:Mreg} that 
$\Mreg$ can be represented as the quotient of
the set
(\ref{eq:Mreg-char}) under the
$U(1)$ action given by 
(\ref{eq:gauge-transf}) with~(\ref{eq:Mreg-uone}). Within~$\mathcal{M}_k$, 
each $U(1)$ equivalence class in (\ref{eq:Mreg-char}) has a 
unique representative that satisfies $p_k=0$ and $u_k>0$. Performing on 
this representative a gauge transformation
(\ref{eq:gauge-transf}) with $g = \mathrm{diag}(u_k^{-1}, u_k)$,
we obtain a point in $\Gamma$ satisfying 
\be
&&
\boldsymbol{u}^2 = \boldsymbol{\pi}^2 >0
\ \ , \ \
\boldsymbol{p}^2=\boldsymbol{v}^2 > 0
\ \ ,
\nn
\\
&&
\boldsymbol{u}\cdot\boldsymbol{p} 
= \boldsymbol{v}\cdot\boldsymbol{\pi} = 0
\ \ ,
\nn
\\
&&
p_k=0
\ \ , \ \
u_k = 1
\ \ .
\label{eq:gauge-Mi}
\ee
It follows that $\mathcal{M}_k$ can be represented as 
the subset of
$\Gamma$ satisfying~(\ref{eq:gauge-Mi}). 

Let now $\tilde{a}, \tilde{b} \in\mathcal{M}_k$
such that $A(\tilde{a}) =
A(\tilde{b})$ for all $A \in \Aclass$. 
Let $a = (\boldsymbol{u},\boldsymbol{p},\boldsymbol{v},\boldsymbol{\pi})$
and $b = (\boldsymbol{u}',\boldsymbol{p}',\boldsymbol{v}',\boldsymbol{\pi}')$ be the
respective representatives of $\tilde{a}$ and $\tilde{b}$
in the gauge~(\ref{eq:gauge-Mi}). We shall
show that $a=b$. We use the basis (\ref{eq:obser}) of~$\Aclass$. 

Consider the observables $A_{ij}$. {}From
$A_{ij}(\tilde{a}) =
A_{ij}(\tilde{b})$ we obtain
\begin{equation}
u_i p_j - u_j p_i = u'_i p'_j - u'_j p'_i
\ \ ,
\label{eq:AA'raw}
\end{equation}
where $1 \le i \le p$ and $1 \le j \le p$.
With $i=k$ and $j\ne k$, the gauge conditions (\ref{eq:gauge-Mi}) show
that (\ref{eq:AA'raw}) reduces to $p_j = p'_j$. The gauge conditions
(\ref{eq:gauge-Mi}) imply directly that $p_k = p'_k$. Hence
$\boldsymbol{p} = \boldsymbol{p}'$.

Multiplying (\ref{eq:AA'raw}) by $p_j$ and summing over $j$ gives
\begin{equation}
\boldsymbol{p}^2 u_i
- (\boldsymbol{u} \cdot\boldsymbol{p}) p_i
=
(\boldsymbol{p} \cdot \boldsymbol{p}') u'_i
-(\boldsymbol{u}'\cdot\boldsymbol{p}) p'_i
\ \ .
\label{eq:AA'}
\end{equation}
Using $\boldsymbol{p} = \boldsymbol{p}'$ 
and~(\ref{eq:gauge-Mi}), (\ref{eq:AA'})~reduces to $u_i =
u'_i$. Hence $\boldsymbol{u} = \boldsymbol{u}'$.

Consider then the observables~$C_{ij}$.
{}From
$C_{ij}(\tilde{a}) =
C_{ij}(\tilde{b})$ we obtain
\begin{equation}
u_i v_j - p_i \pi_j = u'_i v'_j - p'_i \pi'_j
\ \ ,
\label{eq:CC'raw}
\end{equation}
where $1 \le i \le p$ and $1 \le j \le q$. With
$i=k$, (\ref{eq:gauge-Mi}) shows
that (\ref{eq:CC'raw}) reduces to $v_j = v'_j$. Hence
$\boldsymbol{v} = \boldsymbol{v}'$. 

Substituting $\boldsymbol{u} = \boldsymbol{u}'$,
$\boldsymbol{p}
= \boldsymbol{p}'$ and $\boldsymbol{v} = 
\boldsymbol{v}'$ in (\ref{eq:CC'raw}) gives
$p_i(\pi_j - \pi'_j) = 0$. As $\boldsymbol{p}^2 >0$, this implies
$\pi_j = \pi'_j$. Hence $\boldsymbol{\pi} = \boldsymbol{\pi}'$. 
$\blacksquare$

\section{Appendix: $\Aphystar$ on $\Vphys$ for $p=q=3$}
\label{app:pq33-algebra}

In this appendix we analyse the representation of $\Aphystar$
on~$\Vphys$ for $p=q=3$, 
displayed in Table~\ref{table:observables}. We show first
that this representation is irreducible. We then show that the
only inner products in which the
star-relations (\ref{eq:pq33-star}) become adjoint 
relations are multiples of~(\ref{eq:pq33-ip}).


\begin{proposition}
\label{proposition:irred}
Let $U\subset \Vphys$ be a linear subspace invariant under~$\Aphystar$, 
$U \ne \{0\}$. Then
$U = \Vphys$.
\end{proposition}


\myproof
Recall that the operator $\widehat{L^2} := \hat{L}_0^2 + \frac12
\bigl( \hat{L}_+ \hat{L}_- + \hat{L}_- \hat{L}_+ \bigr)$ satisfies 
$\widehat{L^2} \Psi_{lmn} = l(l+1) \Psi_{lmn}$. Let 
$u \in U$, $u\ne0$. Then 
$u = \sum a_{lmn}\Psi_{lmn}$, where only finitely many $a_{lmn}$ are
nonzero. Let $l_0$ be the largest $l$ for which some $a_{lmn}$ is
nonzero. Then 
$u^{(1)} := \prod_{l < l_0} \bigl[ \widehat{L^2} - l(l+1) \bigr] u = 
k \sum_{mn} a_{l_0mn}\Psi_{l_0mn}$, where $k\ne0$. Acting on $u^{(1)}$
finitely many times
with $\hat{L}_+$ and $\hat{J}_+$ gives the vector 
$u^{(2)} = a^{(2)}\Psi_{l_0 l_0 l_0} \ne0$, and 
$u^{(3)} := (\hat{L}_-)^{l_0} (\hat{J}_-)^{l_0} u^{(2)}
= a^{(3)}\Psi_{l_0 00} \ne0$. 
Hence $\Psi_{l_0 00} \in U$.

A direct computation from Table
\ref{table:observables}
shows that
$\hat{J}_- \hat{C}_1^+ \Psi_{l00} - (l-1) \hat{C}_0\Psi_{l00}$ is a
nonzero multiple of
$\Psi_{l+1,00}$ for all $l$ and
$\hat{J}_- \hat{C}_1^+ \Psi_{l00} + (l+1) \hat{C}_0\Psi_{l00}$ is
a nonzero multiple of $\Psi_{l-1,00}$ for $l>0$.
It follows by induction that 
$\Psi_{l00} \in U$ for all~$l$. Acting on
$\Psi_{l00}$ with $\hat{L}_\pm$ and $\hat{J}_\pm$ shows that
$\Psi_{lmn} \in U$ for all values of the indices.
$\blacksquare$


\begin{proposition}
Let $(\cdot \, , \cdot)$ be an inner product in which the
star-relations (\ref{eq:pq33-star})
become adjoint 
relations. Then
$(\Psi_{lmn} \, , \Psi_{l'm'n'}) =
r (2l+1) \delta_{ll'} \delta_{mm'}\delta_{nn'}$,
where $r$ is a positive constant.
\end{proposition}

\myproof
The adjointness relations imply that the 
operator $\widehat{L^2}$ introduced in the proof of Proposition
\ref{proposition:irred} is self-adjoint. Hence 
$l'(l'+1) (\Psi_{lmn},\Psi_{l'm'n'})
= 
(\Psi_{lmn},\widehat{L^2} \Psi_{l'm'n'})
= 
(\widehat{L^2} \Psi_{lmn}, \Psi_{l'm'n'})
= l(l+1) (\Psi_{lmn},\Psi_{l'm'n'})$, which shows that 
$(\Psi_{lmn},\Psi_{l'm'n'})$ vanishes for $l \ne l'$. 
By standard angular momentum techniques 
in the $\mathfrak{o}(3)$ 
subalgebras generated respectively by the 
$\hat{L}$'s and the
$\hat{J}$'s (see for example~\cite{Hecht}), 
we then find 
\begin{equation}
(\Psi_{lmn} \, , \Psi_{l'm'n'}) =
A_l \delta_{ll'} \delta_{mm'}\delta_{nn'}
\ \ ,
\label{eq:app-orthog}
\end{equation}
where $A_l$ depends only on~$l$.

To determine~$A_l$, we use the self-adjointness
of~$\hat{C}_0$. Writing $\Psi_l := \Psi_{l00}$ and using the action of
$\hat{C}_0$ from Table \ref{table:observables}
and~(\ref{eq:app-orthog}), we compute
\be
\frac{(l+1)^2}{2l+1} A_l
&=&
\frac{(l+1)^2}{2l+1} (\Psi_l\ , \Psi_l)
\nn
\\
&=&
(\Psi_l\ , \hat{C}_0 \Psi_{l+1})
\nn
\\
&=&
(\hat{C}_0 \Psi_l\ , \Psi_{l+1})
\nn
\\
&=&
\frac{(l+1)^2}{2l+3}
(\Psi_{l+1}\ , \Psi_{l+1})
\nn
\\
&=&
\frac{(l+1)^2}{2l+3}
A_{l+1}
\ \ , 
\ee
from which by induction $A_l = (2l+1)A_0$.
$\blacksquare$

\section{Appendix: Convergence of the group averaging}
\label{app:convergence}

In this appendix we provide the group averaging
convergence results needed in the main text. 
When not mentioned otherwise, $p$ and $q$ are arbitrary positive
integers. 

To begin, consider $U(g) \Psi_{ljmnk_uk_v}$.
Writing $g$ in the
Iwasawa decomposition~(\ref{eq:iwasawa-decomp1}),
(\ref{eq:Iwasawa}) gives
\begin{equation}
U(g) =
\exp (-i \mu \hat{H}_2 )
\exp ( -i \lambda \hat{D} )
\exp \bigl(-i \theta ( \hat{H}_1 - \hat{H}_2) \bigr)
\ \ .
\label{app-rep-iwasawa}
\end{equation}
As $\Psi_{ljmnk_uk_v}$ is an eigenstate of
$\hat{H}_1 - \hat{H}_2$
with eigenvalue $E_u - E_v$,
(\ref{eq:iwaop-two-action}) yields
\be
U(g) \Psi_{ljmnk_uk_v}
&=&
\frac{z^{(\tilde{j}-\tilde{l})/2}
e^{-i \theta (E_u - E_v)}}
{(2\pi i\mu)^{q/2}}
Y_{lk_u}
\bigl({\theta}^{(u)}\bigr)
\int_0^\infty
dv' \,
u^l {(v')}^{j+q-1}
L_m^{\tilde{l}}(u^2/z)L_n^{\tilde{j}} \bigl( z {(v')}^2 \bigr)
\nn
\\
&&
\times
\exp\left[
- \frac{1}{2}
\left(
\frac{u^2}{z}
+
z {(v')}^2
\right)
+ \frac{i}{2}
\left( \mu u^2 + \frac{ v^2 + {(v')}^2}{\mu} \right)
\right]
\nn
\\
&&
\qquad
\times
\int d\Omega_{v'}\,
\exp\left(- \frac{i}{\mu}(\boldsymbol{v}\cdot\boldsymbol{v}')\right)
Y_{jk_v}
\bigl({\theta}^{(v')}\bigr)
\ \ ,
\label{eq:UPsi-raw1}
\ee
where $z := e^{2\lambda}$ and we are assuming $\mu\ne0$,
$\boldsymbol{v}\ne\boldsymbol{0}$ and $\boldsymbol{u}\ne\boldsymbol{0}$.

We need to evaluate the angular integral
in~(\ref{eq:UPsi-raw1}).
Suppose $q>2$.
We write
$\boldsymbol{v}\cdot\boldsymbol{v}' = v v'
\cos\gamma$
and
expand the
exponential under the angular integral
by (\cite{Vil}, page~98)
\be
e^{it \cos\gamma}=
\frac12 \Gamma\left(\frac{q-2}{2} \right)
\sum_{a=0}^{\infty}{i^a \left({2a+q-2}\right)
\frac{J_{(q-2+2a)/2}(t)}{(t/2)^{(q-2)/2}}
C^{(q-2)/2}_{a}(\cos\gamma)}
\ \ .
\label{eq:planewave}
\ee
We then
expand the Gegenbauer polynomial
$C^{(q-2)/2}_{a}(\cos\gamma)$ as
\be
C^{(q-2)/2}_{a}(\cos\gamma)
=\frac{4\pi^{q/2}}{\Gamma\bigl((q-2)/2\bigr)(2a+q-2)}
\sum_{k}
Y_{a k}
\bigl({\theta}^{(v)}\bigr)
\overline{Y_{a k}
\bigl({\theta}^{(v')}\bigr)}
\ \ , 
\label{eq:Gegen}
\ee
which follows from formula 11.4(2) in \cite{Bate} (correcting a
typographical error in the normalisation factor, as seen from the final
step of the proof on page 247). 
Using the orthonormality of the spherical harmonics, we obtain
\be
&&
\int d\Omega_{v'}\,
\exp\left(-\frac{i}{\mu}(\boldsymbol{v}\cdot\boldsymbol{v}')\right)
Y_{jk_v}
\bigl({\theta}^{(v')}\bigr)
\nn
\\
&&
=
(2 \pi)^{q/2}
i^{-j}
\left(\frac{v v'}{\mu}\right)^{(2-q)/2}
J_{(q-2+2j)/2} (vv'/\mu)
Y_{jk_v}
\bigl({\theta}^{(v)}\bigr)
\ \ .
\label{eq:angular-final}
\ee
For $q=2$, (\ref{eq:angular-final}) follows by
recognising the angular integral as a representation of~$J_j$,
and for $q=1$ it follows from the relation of
$J_{\pm 1/2}$ to trigonometric functions
(\cite{Bate}, Sections 7.3.1 and~7.11).
Hence, for all $p\ge1$ and $q\ge1$, we have
\be
U(g) \Psi_{ljmnk_uk_v}
&=&
\frac{i^{-\tilde{j} - 1} z^{(\tilde{j}-\tilde{l})/2}
e^{-i \theta (E_u - E_v)}}
{\mu}
Y_{lk_u}
\bigl({\theta}^{(u)}\bigr)
Y_{jk_v}
\bigl({\theta}^{(v)}\bigr)
\,
u^{(2-p)/2}
v^{(2-q)/2}
\nn
\\
&&
\times
\int_0^\infty
dv' \,
u^{\tilde{l}}
{(v')}^{\tilde{j}+1}
J_{\tilde{j}} (vv'/\mu)
L_m^{\tilde{l}}(u^2/z)L_n^{\tilde{j}}\bigl(z{(v')}^2\bigr)
\nn
\\
&&
\qquad
\times
\exp\left[
- \frac{1}{2}
\left(
\frac{u^2}{z}
+
z {(v')}^2
\right)
+ \frac{i}{2}
\left( \mu u^2 + \frac{ v^2 + {(v')}^2}{\mu} \right)
\right]
\ .
\phantom{aaa}
\label{eq:UPsi-raw2}
\ee
Performing the integral in (\ref{eq:UPsi-raw2}) gives
(\cite{GradRhyz}, formula 7.421.4)
\be
U(g) \Psi_{ljmnk_uk_v}
&=&
e^{-i \theta (E_u - E_v)}
z^{(\tilde{j}-\tilde{l})/2}
(1+ i\mu z)^{-\tilde{j} - 1}
\left(\frac{1- i\mu z}{1+ i\mu z}\right)^n
Y_{lk_u}
\bigl({\theta}^{(u)}\bigr)
Y_{jk_v}
\bigl({\theta}^{(v)}\bigr)
\nn
\\
&&
\times
\,
u^l v^j
\,
L_m^{\tilde{l}}(u^2/z)
\,
L_n^{\tilde{j}} \! \left(\frac{z v^2}{1+ \mu^2 z^2}\right)
\nn
\\
&&
\times
\exp\left[
- \frac{1}{2}
\left(
\frac{1}{z} - i \mu
\right)
\!
u^2
-
\frac12
\left(
\frac{z}{1 + i \mu z}
\right)
\!
v^2
\right]
\ .
\phantom{aaa}
\label{eq:UPsi-explicit}
\ee

We can now use (\ref{eq:UPsi-explicit}) to prove the convergence
results.

\begin{proposition}
\label{prop:convergence}
Let $\tilde{l}+\tilde{j}>0$. Then
$\bigl( \Psi_{l'j' m' n' k'_uk'_v} ,
U(g)
\Psi_{ljmnk_uk_v}
\bigr)_\mathrm{aux}$
is integrable in absolute value over~$G$.
\end{proposition}

\myproof
It suffices to consider $l'=l$, $j'=j$,
$k'_u = k_u$
and
$k'_v = k_v$,
for otherwise the integrand vanishes. 

In $\overline{\Psi_{ljm'n'k_uk_v}} U(g)
\Psi_{ljmnk_uk_v}$, we use (\ref{eq:testfun})
and (\ref{eq:UPsi-explicit}) and expand the product of the
generalised Laguerre
polynomials as a sum of numerical constants times terms of the form
\begin{equation}
(u^2)^{r'}
(v^2)^{s'}
\left(\frac{u^2}{z}\right)^r
\left(\frac{z v^2}{1+ \mu^2 z^2}\right)^s
\ \ ,
\label{eq:laguerre-expansion-product}
\end{equation}
where $r$, $s$, $r'$ and $s'$ are non-negative integers. Integrating
over $\boldsymbol{u}$ and $\boldsymbol{v}$ term by term, we find that
$\bigl( \Psi_{lj m' n' k_uk_v} ,
U(g)
\Psi_{ljmnk_uk_v}
\bigr)_\mathrm{aux}$
is a sum of terms whose respective absolute values are numerical
constants times
\begin{equation}
\frac{
z^{(\tilde{l} + \tilde{j})/2 + 1 + s + r'}
{(1 + \mu^2 z^2)}^{(s'-s)/2}}
{{\bigl[(z+1)^2 + \mu^2 z^2 \bigr]}^{1 + (\tilde{l} + \tilde{j} + r +
r' + s + s')/2}}
\ \ .
\label{eq:ga-br-integrand}
\end{equation}
An elementary analysis shows that sufficient conditions for
(\ref{eq:ga-br-integrand}) to be integrable over $G$ in the Haar
measure
$e^{2\lambda}\,d\lambda\,d\mu\,d\theta
= \tfrac12 dz\,d\mu\,d\theta$
are 
\be
&&
\tilde{l} + \tilde{j} + 2r + 2s >0
\ \ ,
\nn
\\
&&
\tilde{l} + \tilde{j} + 1 + r + 2s >0
\ \ ,
\label{eq:ga-br-suffconditions}
\ee
which hold since $\tilde{l} + \tilde{j} >0$ by assumption.
$\blacksquare$

\begin{proposition}
\label{prop:fermionic-zero}
Let $\tilde{l}+\tilde{j}>0$ and $p+q \equiv 1 \pmod2$. Then the value of
the integral in Proposition \ref{prop:convergence} is zero. 
\end{proposition}

\myproof
As $p+q \equiv 1 \pmod2$, 
$G$ is the double cover of $\SLtwor$ and the range
of
$\theta$ in (\ref{app-rep-iwasawa}) is
$\theta$ is $0\le
\theta < 4\pi$. 
By Proposition~\ref{prop:convergence}, we may perform the integral over
$\theta$ first, and the 
$\theta$-dependence in (\ref{eq:UPsi-explicit}) shows that this integral
evaluates to zero. 
$\blacksquare$.

\begin{theorem}
\label{theorem:ga-convergence}
Let $p\ge2$, $q\ge2$ and $p+q>4$. Then
the integral in (\ref{eq:ga})
converges in absolute value for all
$\phi_1,\phi_2 \in \Phi$. If $p+q \equiv 1 \pmod2$, the value of the
integral is zero. 
\end{theorem}

\myproof
It suffices to consider
$\phi_1,\phi_2 \in \{
\Psi_{ljmnk_uk_v} \}$. The inequalities on $p$ and $q$
imply that the conditions of Propositions \ref{prop:convergence} and 
\ref{prop:fermionic-zero} are satisfied. 
$\blacksquare$

\begin{proposition}
\label{proposition:fubini}
Let $\tilde{l}>0$ and $\tilde{j}>0$. Then
$\overline{\Psi_{ljm'n'k_uk_v}}
U(g) \Psi_{ljmnk_uk_v}$
is integrable in absolute value over $G \times \BbbR^{p+q}$.
\end{proposition}

\myproof
In $\overline{\Psi_{ljm'n'k_uk_v}} U(g)
\Psi_{ljmnk_uk_v}$, we use (\ref{eq:testfun})
and (\ref{eq:UPsi-explicit}), expand the product of the
generalised Laguerre polynomials as in the proof of Proposition
\ref{prop:convergence} and consider the individual
terms in this expansion. We now \emph{first\/} take the absolute value
and then integrate. The integrals over~$\theta$,
${\theta}^{(u)}$ and ${\theta}^{(v)}$ are
bounded by constants, the integrals over $u$ and $v$ are convergent
and easily performed, and an elementary analysis shows that the
remaining $\int dz \, d\mu$ integral is convergent provided
$\tilde{l}>0$ and $\tilde{j}>0$. 
$\blacksquare$

\begin{theorem}
\label{theorem:ga-13-convergence}
Let $p=1$, $q=3$ and let $\Phi^\mathrm{mod}$ 
be as in
subsection~\ref{subsec:p=1,q=3}.
Then the integral in (\ref{eq:ga})
converges in absolute value for all
$\phi_1,\phi_2 \in \Phi^\mathrm{mod}$.
\end{theorem}

\myproof
The only case not
covered by Proposition \ref{prop:convergence} is $\phi_1 = \psi_{mn}$,
$\phi_2 = \psi_{m'n'}$. 

In $\overline{ \psi_{m'n'} } U(g) \psi_{mn}$, 
we use~(\ref{eq:testfun-11}),
(\ref{eq:psimn-def}) and (\ref{eq:UPsi-explicit}) and expand the 
generalised Laguerre polynomials of argument $u^2/z$ and
$z v^2/(1+ \mu^2 z^2)$ as polynomials in their respective
arguments. Inequalities (\ref{eq:ga-br-suffconditions}) in the
proof of Proposition
\ref{prop:convergence} show that it suffices to keep only the constant
terms of these polynomials. Doing this, and
integrating over $u_1$ and $\boldsymbol{v}$ 
by 7.414.8 in~\cite{GradRhyz}, we
obtain two terms whose absolute values are numerical constants times 
\begin{equation}
\frac{z^{2}}
{\bigl[(z+1)^2 + \mu^2 z^2 \bigr]^2}
\times
\left[
\frac{(z-1)^2 + \mu^2 z^2}
{(z+1)^2 + \mu^2 z^2}
\right]^{(m'+n')/2}
\ \ ,
\end{equation}
which is integrable in the measure $dz \, d\mu$.
$\blacksquare$




\newpage

\begin{thebibliography}{99}



\bibitem{AH}  
A. Higuchi,
\emph{Quantum linearization instabilities of de~Sitter space-time~2\/},
Class.\ Quantum Grav.\ {\bf 8}, 1983
(1991). 

\bibitem{KL}
N.~P. Landsman,
\emph{Rieffel induction as generalized quantum
Marsden-Weinstein reduction\/},
J.~Geom.\ Phys.\ {\bf 15}, 285
(1995). 
\lanln{hep-th/9305088} 

\bibitem{QORD} 
D. Marolf, 
\emph{Quantum observables and recollapsing dynamics\/},
Class.\ Quantum Grav.\ {\bf 12}, 1199
(1995). 
\lanln{gr-qc/9404053}

\bibitem{epistle}
A.~Ashtekar, 
J.~Lewandowski, 
D.~Marolf, 
J.~Mour\~ao, and 
T.~Thiemann,
\emph{Quantization of diffeomorphism invariant
theories of connections with local degrees of
freedom\/},
J.~Math.\ Phys.\ {\bf 36}, 6456
(1995). 
\lanln{gr-qc/9504018}

\bibitem{BC} 
D. Marolf, 
\emph{Refined algebraic quantization: Systems with
a single constraint\/},
in:
{\it Symplectic Singularities and Geometry of Gauge Fields\/} 
(Banach Center Publications, Polish Academy of Sciences, Institute
of Mathematics, Warsaw, 1997). 
\lanln{gr-qc/9508015}  

\bibitem{lands-against} 
N.~P. Landsman, 
\emph{Against the Wheeler-DeWitt equation\/}, 
Class.\ Quantum Grav.\ {\bf 12}, L119
(1995). 
\lanln{gr-qc/9510033}

\bibitem{lands-wren} 
N.~P. Landsman and 
K.~K. Wren, 
\emph{Constrained quantization and $\theta$-angles\/}, 
Nucl.\ Phys.\ {\bf B502}, 537 
(1997). 
\lanln{hep-th/9706178}

\bibitem{GM2}
D.~Giulini and D.~Marolf,
\emph{A uniqueness theorem for constraint quantization\/},
Class.\ Quantum Grav.\ {\bf 16}, 2489
(1999). 
\lanln{gr-qc/9902045}

\bibitem{GoMa} 
A.~Gomberoff 
and 
D.~Marolf,
\emph{On group averaging for $\mathrm{SO}(n,1)$}, 
Int.\ J. Mod.\ Phys.\ D {\bf 8}, 519
(1999). 
\lanln{gr-qc/9902069}

\bibitem{LouRov}
J.~Louko and C.~Rovelli,
\emph{Refined algebraic quantization in the
oscillator representation of $\SLtwor$},
J.\ Math.\ Phys.\ {\bf 41}, 132
(2000).
\lanln{gr-qc/9907004}

\bibitem{Giulini-rev}
D.~Giulini, 
\emph{Group averaging and refined algebraic quantization\/}, 
Nucl.\ Phys.\ Proc.\ Suppl.\  {\bf 88},  385 (2000). 
\lanln{gr-qc/0003040} 

\bibitem{Marolf-MG}
D.~Marolf, 
\emph{Group averaging and refined algebraic quantization: 
Where are we
now?\/}, 
in: 
{\it The Ninth Marcel Grossmann meeting : Proceedings\/}, 
edited by  
V.~G. Gurzadyan, R.~T. Jantzen and R.~Ruffini 
(World Scientific, Singapore, 2002). 
\lanln{gr-qc/0011112}

\bibitem{Shvedov}
O.~Yu.~Shvedov, 
\emph{On correspondence of BRST-BFV, Dirac and refined 
algebraic quantisations of constrained systems\/}, 
Ann.\ Phys.\ (N.Y.) {\bf 302}, 2
(2002). 
\lanln{hep-th/0111270}

\bibitem{GM1}
D.~Giulini and D.~Marolf,
\emph{On the generality of refined algebraic quantization\/},
Class.\ Quantum Grav.\ {\bf 16}, 2479
(1999).
\lanln{gr-qc/9812024}

\bibitem{DeWitt}
B.~S. DeWitt, 
\emph{Quantum theory of gravity.\ 
I. The canonical theory\/}, 
Phys.\ Rev.\ {\bf 160}, 1113
(1967). 

\bibitem{MRT}
M.~Montesinos, C.~Rovelli and T.~Thiemann,
\emph{$\SLtwor$ model with two Hamiltonian constraints\/},
Phys.\ Rev.\ D {\bf 60}, 044009 (1999).
\lanln{gr-qc/9901073}

\bibitem{Monte}
M.~Montesinos, 
\emph{Relational evolution of the degrees of freedom of generally
covariant quantum theories\/},
Gen.\ Rel.\ Grav.\ {\bf 33}, 1
(2001). 
\lanln{gr-qc/0002023}

\bibitem{GamPor}
R.~Gambini
and 
R.~A. Porto,  
\emph{Relational time in generally covariant quantum systems: 
Four models\/},
Phys.\ Rev.\ D {\bf 63}, 105014
(2001). 
\lanln{gr-qc/0101057}

\bibitem{Ash1}
A.~Ashtekar,
\emph{Lectures on Non-Perturbative Canonical Gravity\/}
(World Scientific, Singapore, 1991).

\bibitem{Ash2}
A.~Ashtekar and R.~S.~Tate,
\emph{An algebraic extension of Dirac quantization: Examples\/},
J. Math.\ Phys.\ {\bf 35}, 6434
(1994).
\lanln{gr-qc/9405073}

\bibitem{TruSL}
M.~Trunk,
\emph{An $\SLtwor$ model of constrained systems:
Algebraic constraint quantization\/},
University of Freiburg preprint THEP 99/3. 
\lanln{hep-th/9907056}

\bibitem{trunk-kepler} 
M.~Trunk, 
\emph{The five-dimensional Kepler problem as an
$SU(2)$ gauge system: Algebraic constraint quantization\/},
Int.\ J. Mod.\ Phys.\ A {\bf 11}, 2329
(1996). 
\lanln{hep-th/9510019}

\bibitem{trunk-pseudorigid} 
M.~Trunk, 
\emph{Algebraic constraint quantization and the
pseudo rigid body\/}, 
University of Freiburg preprint THEP~96/17. 
\lanln{hep-th/9701112} 

\bibitem{bojo-etal}
M.~Bojowald, 
H.~A. Kastrup, 
F.~Schramm, 
and 
T.~Strobl, 
\emph{Group theoretical quantization of a 
phase space $S^1 \times \BbbR^+$ 
and the mass spectrum of Schwarzschild black holes in $D$
space-time dimensions\/}, 
Phys.\ Rev.\ D {\bf 62}, 044026 (2000). 
\lanln{gr-qc/9906105}

\bibitem{bojo-strobl} 
M.~Bojowald
and 
T.~Strobl, 
\emph{Group theoretical quantization 
and the example of a phase space $S^1 \times \BbbR^+$}, 
J. Math.\ Phys.\ {\bf 41}, 2537
(2000). 
\lanln{quant-ph/9908079}

\bibitem{kastrup} 
H.~A. Kastrup, 
\emph{Quantization of the optical phase space 
${\mathcal S}^2 =\{\varphi \bmod{2\pi}, I >0\}$
in terms of the group
$\mathrm{SO}^{\uparrow}(1,2)$}, 
Fortsch.\ Phys.\ {\bf 51}, 975
(2003). 
\lanln{quant-ph/0307069}

\bibitem{isham-les}
C.~J. Isham,
\emph{Topological and global aspects of quantum theory\/}, 
in: {\it Relativity, Groups and Topology~II: 
Les Houches 1983\/}, 
edited by B.~S. DeWitt and R.~Stora
(North-Holland, Amsterdam, 1984).

\bibitem{GuiSte}
V.~Guillemin 
and 
S.~Sternberg, 
\emph{Symplectic Techniques in Physics\/} 
(Cambridge University Press, Cambridge, 1984)

\bibitem{BarsKounnas}
I.~Bars and C.~Kounnas, 
\emph{Theories with Two Times\/}, 
Phys.\ Lett.\ {\bf B402}, 25
(1997). 
\lanln{hep-th/9703060}

\bibitem{Bars98}
I.~Bars, 
\emph{Two-Time Physics\/}, 
in: 
\emph{Proceedings of the 22nd International Colloquium On Group
Theoretical Methods In Physics\/}, 
edited by S.~P. Corney, R.~Delbourgo
and 
P.~D. Jarvis
(International Press, Cambridge, Massachusetts, 1999), 
2--17. 
\lanln{hep-th/9809034}. 

\bibitem{Bars00}
I.~Bars, 
\emph{Survey of Two-Time Physics\/}, 
Class.\ Quantum Grav.\ {\bf 18}, 3113
(2001). 
\lanln{hep-th/0008164}

\bibitem{Bars01}
I.~Bars, 
\emph{2T-Physics 2001\/}, 
in: 
\emph{New developments in fundamental interaction theories: 
Proceedings of the 37th Karpacz Winter School Of Theoretical
Physics\/} 
(AIP Conference Proceedings, Vol.\ 589), 
edited by 
J.~Lukierski and J.~Rembielinski 
(American Institute of Physics, Melville, NY, 2001), 
18--30. 
\lanln{hep-th/0106021}. 

\bibitem{Dittrich-pc}
B.~Dittrich, private communication (2003). 

\bibitem{Thiemann-master} 
T.~Thiemann, 
\emph{The Phoenix Project: 
Master constraint program for loop quantum gravity\/}, 
\lanln{gr-qc/0305080}. 

\bibitem{SchallerStrobl}
P.~Schaller and T.~Strobl, 
\emph{Diffeomorphisms versus nonabelian gauge transformations: an
example of $(1+1)$ dimensional gravity\/}, 
Phys.\ Lett.\ {\bf B337}, 266
(1994). 
\lanln{hep-th/9401110}

\bibitem{Tuyn}
G.~M. Tuynman,
\emph{Reduction, quantization, and nonunimodular groups\/},
J. Math.\ Phys.\ {\bf 31}, 83
(1990).

\bibitem{DEG}
C.~Duval,
J.~Elhadad
and
G.~M. Tuynman,
\emph{The BRS method and geometric quantization: some examples\/},
Commun.\ Math.\ Phys.\ {\bf 126}, 535
(1990).

\bibitem{Howe}
R. Howe and E.~C. Tan, 
\emph{Non-Abelian Harmonic
Analysis:  Applications of $\SLtwor$} 
(Springer, New York, 1992).

\bibitem{Dir3}
P.~A.~M. Dirac,
\emph{Lectures on Quantum Mechanics\/}
(Belfer Graduate School of Science, New York, 1964).

\bibitem{Hen}
M.~Henneaux and C.~Teitelboim,
\emph{Quantization of Gauge Systems\/}
(Princeton University Press, Princeton, 1992).

\bibitem{Gil}
R.~Gilmore,
\emph{Lie Groups, Lie Algebras and Some of Their Applications\/}
(John Wiley \& Sons, Inc., 1974).

\bibitem{AMMP}
J.~A. Azc\'arraga, A.~J. Macfarlane,
A.~J. Mountain and J.~C. P\'erez Bueno,
\emph{Invariant tensors for simple groups\/},
Nucl.\ Phys.\ B {\bf 510}, 657
(1998).
\lanln{physics/9706006}

\bibitem{Bate}
A.~Erd\'elyi, W.~Magnus, F.~Oberhettinger, and
F.~G. Tricomi,
\emph{Higher Transcendental Functions, Vol. 2\/}
(Bateman Manuscript project,
McGraw-Hill, New York, 1953).

\bibitem{Vil}
N.~Ja.\ Vilenkin and A.~V. Klimyk,
\emph{Representation of Lie Groups and Special Functions\/}, 
Vol.~2
(Kluwer Academic Publishers, 1993).

\bibitem{arfken}
G.~Arfken,
{\it Mathematical Methods for Physicists\/},
second edition
(Academic, New York, 1970).

\bibitem{magnusetal}
W.~Magnus,
F.~Oberhettinger,
and
R.~P. Soni,
{\it Formulas and Theorems for the Special
Functions of Mathematical
Physics}, 3rd ed.\ (Springer, Berlin, 1966).

\bibitem{Howe-dual}
R.~Howe, 
\emph{On some results of Strichartz 
and of Rallis and Schiffman\/}, 
J.~Funct.\ Anal.\ {\bf 32}, 297
(1979).

\bibitem{PaulTan}
A.~Paul and E.~C. Tan,
\emph{On the dual pairs
$\left(\mathrm{O}(p,q), \SLtwor\right)$,
$\left(\mathrm{U}(p,q), \mathrm{U}(1,1)\right)$
and
$\left(\mathrm{Sp}(p,q), \mathrm{O}^*(4)\right)$\/},
Pacif.\ J. Math.\ {\bf 187}, 349 (1999).

\bibitem{Hecht}
K.~T. Hecht,
\emph{Quantum Mechanics\/}
(Springer, New York, 1993), 
Section 14C.

\bibitem{GradRhyz}
I.~S.
Gradshteyn
and
I.~M.
Ryzhik,
{\it Table of Integrals, Series and Products},
4th edition
(Academic, New York, 1980).


\end{thebibliography}
\end{document}